\documentclass[prb,twocolumn,showpacs,superscriptaddress,floatfix]{revtex4}

\usepackage{hyperref}
\usepackage{graphicx}
\usepackage{color}
\usepackage{amssymb,amsmath}

\newcommand{\ic}        { {\rm i}}
\newcommand{\iz}        { \ic 0^{+}}
\newcommand{\order}[1]  { {{\cal O}\left(#1\right)}}
\newcommand{\sorder}[1] { {{\cal O}(#1)}}

\newcommand{\sfrac}[2]  { {\textstyle \frac{#1}{#2}} }

\newcommand{\abs}[1]    { {\left| #1 \right|}}
\newcommand{\mean}[1]   { \left<  \, {#1} \, \right> }
\newcommand{\comm}[2]   { \left[#1,#2\right] }
\newcommand{\GF}[2]     { \left<\!\left< \, {#1}; \, {#2} \, \right>\!\right> }
\newcommand{\real}      { {\rm Re\,} }
\newcommand{\imag}      { {\rm Im\,} }
\newcommand{\sign}[1]   { {\rm sgn}(#1) }

\newcommand{\vect}[1]   { {\mathbf #1 } }
\newcommand{\vk}        { {\vect{k}} }

\newcommand{\up}	{ \uparrow }
\newcommand{\dn}	{ \downarrow }

\newcommand{\Gi}[1]     { G_{#1} }
\newcommand{\Di}[1]	{ D_{#1} }
\newcommand{\Si}[1]	{ \Sigma_{#1} }

\newcommand{\GiF}	{ g_{} }

\newcommand{\GiRHF}[1]	{ \Gi{#1}^{\rm RHF} }

\newcommand{\GiUHF}	{ \Gi{}^{\rm UHF} }
\newcommand{\DiUHF}	{ \Di{}^{\rm UHF} }

\newcommand{\Guhf}[2]   { {\cal G}_{#1 #2} }
\newcommand{\GAuhf}[1]  { \Guhf{\rm A}{#1} }
\newcommand{\GBuhf}[1]  { \Guhf{\rm B}{#1} }
\newcommand{\Gauhf}[1]  { \Guhf{}{#1} }

\newcommand{\GAsuhf}	{ \GAuhf{\sigma} }

\newcommand{\Gasuhf}	{ \Gauhf{\sigma} }

\newcommand{\Duhf}[2]   { D^{0}_{#1 #2} }
\newcommand{\DAuhf}[1]  { \Duhf{\rm A}{#1} }
\newcommand{\DBuhf}[1]  { \Duhf{\rm B}{#1} }
\newcommand{\Dauhf}[1]  { \Duhf{}{#1} }

\newcommand{\DAsuhf}	{ \DAuhf{\sigma} }

\newcommand{\Dasuhf}	{ \Dauhf{\sigma} }

\newcommand{\GA}[1]	{ G_{{\rm A} #1} }
\newcommand{\GB}[1]	{ G_{{\rm B} #1} }
\newcommand{\Ga}[1]	{ G_{#1} }
\newcommand{\Gal}[1]	{ G_{\alpha #1} }
\newcommand{\Gas}	{ G_{\sigma} }

\newcommand{\ReDelta}   { \Delta_{{\rm R}} }
\newcommand{\ImDelta}   { \Delta_{{\rm I}} }
\newcommand{\Do}	{ \Delta_{0} }

\newcommand{\Sa}[1]	        { \Sigma_{#1} }
\newcommand{\fullSa}[1]	        { \tilde{\Sigma}_{#1} }

\newcommand{\ReSa}[1]		{ \Sa{#1}^{\rm R} }
\newcommand{\ImSa}[1]		{ \Sa{#1}^{\rm I} }
\newcommand{\RefullSa}[1]	{ \fullSa{#1}^{\rm R} }
\newcommand{\ImfullSa}[1]	{ \fullSa{#1}^{\rm I} }

\newcommand{\Pipm}[1]	{ \Pi^{+-}_{#1 #1} }
\newcommand{\Pimp}[1]	{ \Pi^{-+}_{#1 #1} }
\newcommand{\Piapm}   	{ \Pipm{} }
\newcommand{\Piamp}   	{ \Pimp{} }

\newcommand{\Popm}[1]	{ ^{0}\Pi^{+-}_{#1 #1} }

\newcommand{\Poapm}   	{ \Popm{} }

\newcommand{\w}		{ \omega }
\newcommand{\wo}	{ \omega_{0} }
\newcommand{\wtil}      { \tilde{\w} }
\newcommand{\wm}	{ \omega_{\rm m} }
\newcommand{\qm}	{ q_{\rm m} }

\newcommand{\epsi}	{ \varepsilon_{i} }
\newcommand{\epsis}	{ \varepsilon_{i\sigma} }
\newcommand{\mi}	{ \mu }
\newcommand{\mo}	{ \mu_{0} }

\newcommand{\Uc}	{ U^{{\rm c}} }
\newcommand{\Uco}	{ U_{0}^{{\rm c}} }

\newlength{\Units}
\newlength{\FigureWidth}

\begin{document}

\setlength{\FigureWidth}{0.95\columnwidth} 

\title{The Anderson impurity model with a narrow-band host: from
  orbital physics to the Kondo effect}

\author{Steffen Sch\"afer} 

\affiliation{Aix-Marseille Universit\'e and Institut Mat\'eriaux
  Micro\'electronique Nanosciences de Provence (IM2NP), Facult\'e des
  Sciences de St J\'er\^ome, 13397 Marseille, France}

\date{September 27, 2010; revised manuscript March 25, 2011}
	
\begin{abstract}

  A particle-hole symmetric Anderson impurity model with a metallic
  host of narrow bandwidth is studied within the framework of the
  local moment approach.  The resultant single-particle spectra are
  compared to unrestricted Hartree-Fock, second order perturbation theory
  about the noninteracting limit, and Lanczos spectra by Hofstetter
  and Kehrein. Rather accurate analytical results explain the spectral
  evolution over almost the entire range of interactions. These
  encompass, in particular, a rationale for the four-peak structure
  observed in the low-energy sector of the Lanczos spectra in the
  moderate-coupling regime. In weak coupling, the spectral evolution
  is governed by orbital effects, while in the strong coupling Kondo
  limit, the model is shown to connect smoothly to the generic
  Anderson impurity with a flat and infinitely wide hybridization
  band.

\end{abstract}

% %%%%%%%%%%%%%%%%%%%%%%%%%%%%%%%%%%%%%%%%%%%%%%%%%%%%%%%%%%%%
% PACS numbers according to 2010 PACS
% %%%%%%%%%%%%%%%%%%%%%%%%%%%%%%%%%%%%%%%%%%%%%%%%%%%%%%%%%%%%

% other possibilities are:
% 71.30.+h. % Metal-insulator transitions and other electronic transitions.  

\maketitle

% %%%%%%%%%%%%%%%%%%%%%%%%%%%%%%%%%%%%%%%%%%%%%%%%%%%%%%%%%%%%
\section{Introduction}

Five decades of intense experimental and theoretical research have
boosted the Anderson impurity model (AIM) \cite{Anderson61} far beyond
the scope of the ``localized magnetic states in metals'' that it was
initially designed for.  In the first three decades, its concept of a
single level with on-site Coulomb repulsion coupled to a host without
electronic interactions was mainly used to describe magnetic
transition metal impurities dissolved in otherwise nonmagnetic bulk
metals. The last two decades' extraordinary progress in
nanotechnology, however, brought a myriad of new and rather surprising
implementations of what in the meantime had become one of the
theorists' favourite toys, ranging from tunable quantum dots
\cite{Sasaki00,vanderWiel00} over carbon nanotubes
\cite{Odom00,Nygard00} and adsorbed organic molecules \cite{Zhao05} to
single-electron \cite{GoldhaberGordon98} or single-molecule
transistors.  \cite{Liang02}

The vast majority of theoretical work, comprehensively reviewed in
Ref.~\onlinecite{Hewson}, focuses on AIMs with metallic hosts of large
bandwidth. In the limit of infinite bandwidth, the exact static and
thermodynamic properties can be deduced from the Bethe ansatz
solution.  \cite{Tsvelik83,Okiji83,Schlottmann89} A powerful and
versatile alternative is provided by the numerical renormalization
group (NRG), \cite{Krish80a,Krish80b} recent extensions of which have
also been able to address the dynamics of the model to excellent
accuracy. \cite{Frota86,Sakai89,Costi90,Hofstetter00,Dickens01,Galpin05,Bulla08}

Yet another field of application is the Mott metal-to-insulator
transition in high spatial dimensions:
\cite{MetznerVollhardt89,Georges96,Gebhard} here, dynamical mean-field
theory (DMFT) \cite{Georges96} reduces the at first sight unrelated
problem of interacting electrons on a high-dimensional lattice to an
effective AIM immersed in a bath of identical sites whose properties
have to be determined self-consistently. It was within this context
that the possibility of an AIM with a metallic host of narrow
bandwidth was first evoked \cite{Hofstetter99} since it naturally
arises, in the vicinity of the Mott-Hubbard transition,
\cite{Kehrein98} within the now widely accepted scenario of a metallic
state surrounded by a preformed gap. \cite{Georges96} At the time,
W.~Hofstetter and S.~Kehrein argued, on grounds of their
Lanczos-determined single-particle spectra for a corresponding AIM,
\cite{Hofstetter99} that this incipient gap might be populated by
localized states. Although these states have not actually been
observed {\em inside} the preformed gap of the infinite-dimensional
Hubbard model, recent high-resolution dynamic density-matrix
renormalization group (DDMRG) data
\cite{Nishimoto04,Karski05,Karski08} do indeed show very narrow
features on the inner band edges of the Hubbard satellites in the
appropriate regime of interactions.

Independent from the relevance for the Mott-Hubbard transition,
several questions about these sharp features may arise: (i) what are
the underlying physical processes? (ii) do these processes depend on
correlations within the DMFT bath?  (iii) why do these features
disappear for both, small and large values of the Coulomb repulsion?
(iv) are they related -- and if yes, in which manner -- to the series
of peaks observed in the low-energy sector of Lanczos-determined
spectra of an AIM with a correlationless narrow-band host?
\cite{Hofstetter99} and (v) why, in the latter case, are the peaks
organized in a four-set structure?

Some of the above questions have already been addressed in a previous
article, \cite{SL01} in which the AIM was mainly studied
perturbatively about the noninteracting limit; others, in particular
those concerning the rich low-energy structure at moderate interaction
strengths or the strong-coupling Kondo regime, lie out of reach for
perturbative approaches and remain unanswered thus far.
It is these questions, among others, that shall be addressed in the
present article, primarily within the framework of the so-called local
moment approach (LMA). This nonperturbative many-body Green function
formalism, developed by D.~Logan and co-workers,
\cite{Logan98,Dickens01,Glossop02} introduces the concept of local
moments from the outset. At pure mean-field level, this would lead to
a doubly degenerate ground-state, as appropriate for an insulator, but
in manifest contradiction to the Kondo singlet observed for impurities
hosted in metals. The LMA aims to transcend this deficiency by
accounting for dynamical tunneling processes between the two
mean-field ground-states, at a rate which has to be determined in
consistency with Fermi-liquid behaviour on the lowest energy
scale. The resulting formalism is equally well adapted to impurities
in metallic \cite{Logan98} and insulating \cite{Galpin08b} hosts, be
it in the particle-hole symmetric limit \cite{Logan98} or away from
it. \cite{Glossop02} The LMA has so far been applied and adapted to a
variety of impurity and lattice problems with strong electronic
correlations, where its ability to cope with essentially all
interaction regimes and to correctly describe all energy scales has
proven very
valuable. \cite{Glossop00,Logan01a,Logan01b,Smith03,Logan05,Galpin09}
In the present case of an impurity hosted in a narrow band, the LMA
has to be generalized by using a renormalized, sum-rule compliant
version of the original ladder-sum propagator for the transverse spin
fluctuations. This extension is necessary, especially in the regime of
moderate interaction strengths, to correctly capture the subtle
low-energy dynamics (see Secs.~\ref{sec:Pi} and
\ref{sec:spectra_medU}).  

The article is outlined as follows: after a brief presentation of the
model, Sec.~\ref{sec:mf} defines the narrow-band regime and presents
two different mean-field solutions. Sec.~\ref{sec:lma} starts with a
review of the LMA, followed by a discussion of the physically relevant
transverse spin fluctuations and of two important sum rules for the
associated polarization propagator; its last paragraph is dedicated to
the self-energy approximation implemented in practice.
Sec.~\ref{sec:spectra} presents the LMA impurity spectra for different
regimes defined by the strength of the on-site Coulomb interaction
$U$; the spectra are compared to corresponding Hartree-Fock or
perturbation theory results and, where available, to Hofstetter and
Kehrein's Lanczos-determined spectra. \cite{Hofstetter99} A Conclusion
section closes the paper.

% %%%%%%%%%%%%%%%%%%%%%%%%%%%%%%%%%%%%%%%%%%%%%%%%%%%%%%%%%%%%
\section{Hamiltonian and mean field theories}
\label{sec:mf}

The Hamiltonian for the AIM is given in standard notation by
\begin{eqnarray}
\label{Hamiltonian}
\hat{H} & = & 
\sum\limits_{\vk\sigma}\varepsilon_{\vk}\hat{n}_{\vk\sigma}
\,+\,
\sum\limits_{\sigma} \epsi\hat{n}_{i\sigma} 
\,+\,
U\hat{n}_{i\up} \hat{n}_{i\dn} 
\nonumber \\ &&
\,+\,
\sum\limits_{\vk\sigma} 
\left( V_{i\vk} c^{+}_{i\sigma} c_{\vk\sigma} + \mathrm{h.c.} \right)
\end{eqnarray}
where the first term describes electrons (of spin $\sigma=\up,\dn$) in
a metallic host band of dispersion $\varepsilon_{\vk}$. The following
two terms refer to the impurity, with $\epsi$ the impurity level and
$U$ the on-site Coulomb interaction. The final term describes the
one-electron hybridization between the impurity and host. 

Throughout this article, as in Refs.~\onlinecite{Hofstetter99} and
\onlinecite{SL01}, the particle-hole symmetric AIM, obtained by
setting $\epsi=-U/2$, will be studied. In this case, the empty and
doubly occupied impurity states are degenerate, whence for all
interaction strengths the Fermi level remains fixed at its
noninteracting value and the impurity charge is
$n_{i}=\mean{\hat{n}_{i\up}+\hat{n}_{i\dn}}=1$.  Regardless of the
interaction strength, single-particle spectra are thus symmetric with
respect to the Fermi level, $\w=0$. Persistent charge fluctuations
guarantee the system's metallic character, allowing for the
possibility to recast the exact single-particle impurity Green
function as an infinite-order perturbation series, with each diagram
depending solely on $U$ and the noninteracting Green function
\begin{equation}
  \label{GFree}
  \GiF(\w) \,=\, \left[\w+\iz\sign{\w}-\Delta(\w)\right]^{-1} \;
  \mbox{.}
\end{equation}
In the latter expression,
$\Delta(\w)=\ReDelta(\w)-\ic\,\sign{\w}\ImDelta(\w)$ stands for the
hybridization function
\begin{equation}
  \label{hybrid}
  \Delta(\w) \,=\, 
  \sum\limits_{\vk} \frac{\abs{V_{i\vk}}^{2}}{\w+\iz\sign{\w}-\varepsilon_{\vk}}
\end{equation}
which condenses all relevant information about the host dispersion
$\varepsilon_{\vk}$ and the hybridization matrix elements
$V_{i\vk}$.  

In this section, the single-particle impurity spectra of the AIM will
be calculated in two different mean-field descriptions: Restricted
Hartree-Fock (RHF), on the one hand, implements spin symmetry from the
outset via identical impurity occupation numbers for both spin
species; unrestricted Hartree Fock (UHF), on the other hand, seeks to
determine the occupation numbers self-consistently --- thus allowing
for solutions with different impurity occupancies for $\up$ and
$\dn$-spins at an intermediate stage --- and restores the full spin
symmetry only at the very end.

In both versions of the theory, the (causal) single-particle impurity
Green function $\Guhf{i}{\sigma}(\w)=
\real\Guhf{i}{\sigma}(\w)-\ic\,\sign{\w}\pi\Duhf{i}{\sigma}(\w)$ can
be deduced {\sl via} standard techniques, e.g., the equation-of-motion
method, after Hartree-Fock factorizing the 2-body term in the
Hamiltonian (\ref{Hamiltonian}), $\hat{n}_{i\up}\hat{n}_{i\dn} \simeq
\hat{n}_{i\up} \mean{\hat{n}_{i\dn}} + \mean{\hat{n}_{i\up}}
\hat{n}_{i\dn} - \mean{\hat{n}_{i\up}} \mean{\hat{n}_{i\dn}}$,
yielding
\begin{equation}
\label{Guhf}
\Guhf{i}{\sigma}(\w)\,=\,
\left[\w+\iz\sign{\w}-\epsis-\Delta(\w)\right]^{-1}
\end{equation}
where $\epsis=\epsi+\sfrac{U}{2} (n_{i}-\sigma\mi)$ denotes the
Hartree-Fock corrected impurity level.  In the particle-hole symmetric
AIM, where $\epsi=-U/2$ and
$n_{i}=\mean{\hat{n}_{i\up}+\hat{n}_{i\dn}}=1$, the Hartree-Fock
corrected impurity level solely depends on the impurity moment
$\mi=\mean{\hat{n}_{i\up}-\hat{n}_{i\dn}}$, viz.
\begin{equation}
\label{epsisig}
\epsis\,=\,-\frac{U}{2} \sigma {\mi}
\end{equation}
with $\sigma=+$ $(-)$ for $\up$ ($\dn$) spins.  As a consequence of
the Hamiltonian's invariance under spin inversion, solutions with a
nonvanishing magnetic moment are doubly degenerate, $\mi=+\abs{\mi}$
and $-\abs{\mi}$.

The present paper focuses on AIMs with narrow metallic host bands,
whose width is much smaller than the hybridization strength at the
Fermi level, $\Do=\ImDelta(\w=0)$.  The physics of such a narrow-band
AIM is naturally rather insensitive to the precise form of the
hybridization, meaning that, without loss of generality,
$\ImDelta(\w)$ may be assumed to consist of a single flat band of
intensity $\Do$, ranging from $-D$ to $+D$, with $D\ll\Do$. As for any
time-ordered Green function the real part follows via Hilbert
transform,
\begin{equation}
\label{Hilbert}
\ReDelta(\w) \,=\, 
{\cal P}\int\limits_{-\infty}^{+\infty} \frac{{\rm d}\w'}{\pi}
\frac{\ImDelta(\w')}{\w-\w'}
\;\mbox{,}
\end{equation}
so that in total
\begin{equation}
\label{nbhybrid}
\Delta(\w) \,=\, 
\frac{\Do}{\pi} \, \ln{\abs{\frac{\w+D}{\w-D}}} 
\,-\,
\ic \Do\,\sign{\w}\,\theta(D-\abs{\w})
\;\mbox{.}
\end{equation}

% %%%%%%%%%%%%%%%%%%%%%%%%%%%%%%%%%%%%%%%%
\subsection{Restricted Hartree-Fock (RHF)}
\label{sec:rhf}

The spin-reversal invariance of the Anderson Hamiltonian
(\ref{Hamiltonian}) implies that the {\em average} number of $\up$ and
$\dn$-spin electrons on the impurity has to be the same for {\em any}
interaction strength. RHF theory acknowledges this fact from the
outset by enforcing $\mean{\hat{n}_{i\up}}=\mean{\hat{n}_{i\dn}}$ for
all interactions strengths, thus entailing $\epsis\equiv 0$ in
eq.~(\ref{Guhf}). As a result, RHF recovers the noninteracting Green
function for both spin species and all interaction strengths:
\begin{equation}
\label{Grhf}
\GiRHF{\sigma}(\w) \,\equiv \,
\GiF(\w) \,=\, \left[\w+\iz\sign{\w}-\Delta(\w)\right]^{-1}
\end{equation}

The associated single-particle spectrum $\Duhf{i}{\sigma}(\w)$
consists of two contributions: (i) a continuum for $\w\in[-D,D]$,
arising from the nonzero imaginary part of the hybridization; and
(ii) two poles, one lying above and the other symmetrically below the
continuum. The fraction of spectral weight residing in the poles
depends strongly on the bandwidth of the host metal.  In the usual
wide-band model, $D\gg\Do$, most of the spectral intensity is
concentrated in the single-particle band, while the poles are
exponentially weak and hence irrelevant in practice.

The situation is, however, radically different in the present
narrow-band model, defined by $D\ll\Do$. Here, almost the entire
spectral intensity resides in the poles, occurring at frequencies
$\pm\wo$ far outside the band. By analogy with an ${\rm H}_{2}$
molecule, these poles can be viewed as a bonding and an anti-bonding
orbital, thus suggesting that the narrow host band to which the
impurity is coupled behaves effectively as a single site or level.
\cite{Hofstetter99,Lange98}

In this case, the pole frequencies and weights can be obtained to good
accuracy from eq.~(\ref{Grhf}) by using the expansion
$\ReDelta(\w)\sim(2/\pi)\Do D/\w$, valid for $\abs{\w}\gg D$:
\begin{subequations}
\begin{eqnarray}
\label{omega0}
\wo& \simeq & \sqrt{\sfrac{2}{\pi}\Do D}
\quad\\
\label{weight0}
q  & \simeq &\frac{1}{2}-\frac{8 D}{\pi\Do}
\;\mbox{.}
\end{eqnarray}
\end{subequations}
The ``bonding energy'' $\wo$ corresponds to the {\em integrated}
hybridization, or the total hopping between host and impurity;
\cite{SL01} subject to $D\ll\wo\ll\Do$, it defines a second low-energy
scale relevant in the narrow-band regime.

In the following, it will sometimes be helpful to consider $D$ and
$\wo$ as independent parameters. In particular, in the limit of an
infinitely narrow host band, $D\to 0$, this allows us to treat the
host as a single level which couples via a finite bonding energy $\wo$
to the impurity -- a picture henceforth referred to as the {\em
  two-site approximation}.

In addition to the orbital levels of the two-site approximation, the
full noninteracting single-particle spectrum encompasses the
aforementioned Fermi liquid continuum stemming from the hybridization
band on the lowest energy scale, $\abs{\w}\le D$. Its integrated
weight is of the order $\order{D/\Do}$ and thus weak in the
narrow-band regime.

The RHF description, naturally exact in the noninteracting limit, is
expected to break down if $U$ is much larger than the ``molecular''
bonding energy $\wo$: in this case, the extra electron probed by
$\Gi{\sigma}$ is most likely to be introduced on an already singly
occupied impurity which involves an energy cost of the order of the
interaction strength $U$; the single-particle spectra will then be
dominated by Hubbard poles separated by the Coulomb interaction $U$
rather than the molecular orbitals at $\w=\pm\wo$.

% %%%%%%%%%%%%%%%%%%%%%%%%%%%%%%%%%%%%%%%%
\subsection{Unrestricted Hartree-Fock (UHF)}
\label{sec:uhf}

In UHF theory, the magnetic moment residing on the impurity will be
determined self-consistently from the impurity Green function
$\Guhf{i}{\sigma}$ itself. For small interaction strengths, it is
found to be zero, and UHF recovers the noninteracting solution
$\GiF$.  This nonmagnetic solution becomes unstable above some
critical interaction $\Uco$ --- which turns out to be related to the
``molecular'' bonding energy $\wo$ --- and UHF then converges to a
solution with a finite impurity moment. The Hamiltonian
(\ref{Hamiltonian}) is, however, still invariant under spin
inversions, thus guaranteeing for any mean-field ground state with
positive moment $\mi=+{\mo}$ the existence of another degenerate
ground state with opposite moment, $\mi=-{\mo}$. Subsequently, quantum
mechanical tunneling processes between these mean-field ground states
ensure their occurrence with equal probability:
\begin{equation}
\label{Guhfmix}
\GiUHF(\w)\,=\,
\frac{1}{2}\,\left[\GAuhf{\sigma}(\w)\,+\,\GBuhf{\sigma}(\w)\right]
\end{equation}
Throughout the present paper and without loss of generality, ${\rm
  A}$- (${\rm B}$-) type impurities are assumed to be predominantly
$\up$ ($\dn$) spin occupied, implying $\mo\ge 0$.  The $\up\dn$ spin
symmetry of the Hamiltonian entails
$\GAuhf{\sigma}(\w)=\GBuhf{-\sigma}(\w)$, thus guaranteeing the spin
independence of $\GiUHF(\w)$ in eq.~(\ref{Guhfmix}).

The necessity of the mixing process (\ref{Guhfmix}) becomes
particularly obvious in the atomic limit, defined by vanishing
hybridization matrix elements, $V_{i\vk}\equiv 0$. In this limit, the
impurity propagator is $\Gi{}^{\rm AL}(\w) =
\frac{1}{2}([z+\frac{U}{2}]^{-1}+[z-\frac{U}{2}]^{-1})$ [with
  $z=\w+\iz\sign{\w}$].  Each of the two contributions in
eq.~(\ref{Guhfmix}) yields one term of this exact result [via
  $\Delta(\w)\equiv 0$ and $\mo=1$]: an $\up$-spin electron can only
be retrieved from an ${\rm A}$-type impurity (first term), and can
only be added to a ${\rm B}$-type impurity (second term).

In practical terms, the Green function
$\GAsuhf(\w)=\real\GAsuhf(\w)-\ic\,\sign{\w}\pi\DAsuhf(\w)$ is
obtained from eqs.~(\ref{Guhf}) and (\ref{epsisig}) with an impurity
moment $\mi=+\mo$, calculated self-consistently from
\begin{equation}
\label{scmu}
\mo\,=\,\int\limits_{-\infty}^{0} {\rm d}\w 
\left[ \DAuhf{\up}(\w)- \DAuhf{\dn}(\w)\right]
\;\mbox{;}
\end{equation}
[$\GBuhf{\sigma}(\w)$ follows equivalently for $\mi=-\mo$].  
\begin{figure}[htbp]
  \begin{center} 
  \leavevmode 
  \includegraphics[width=\FigureWidth]{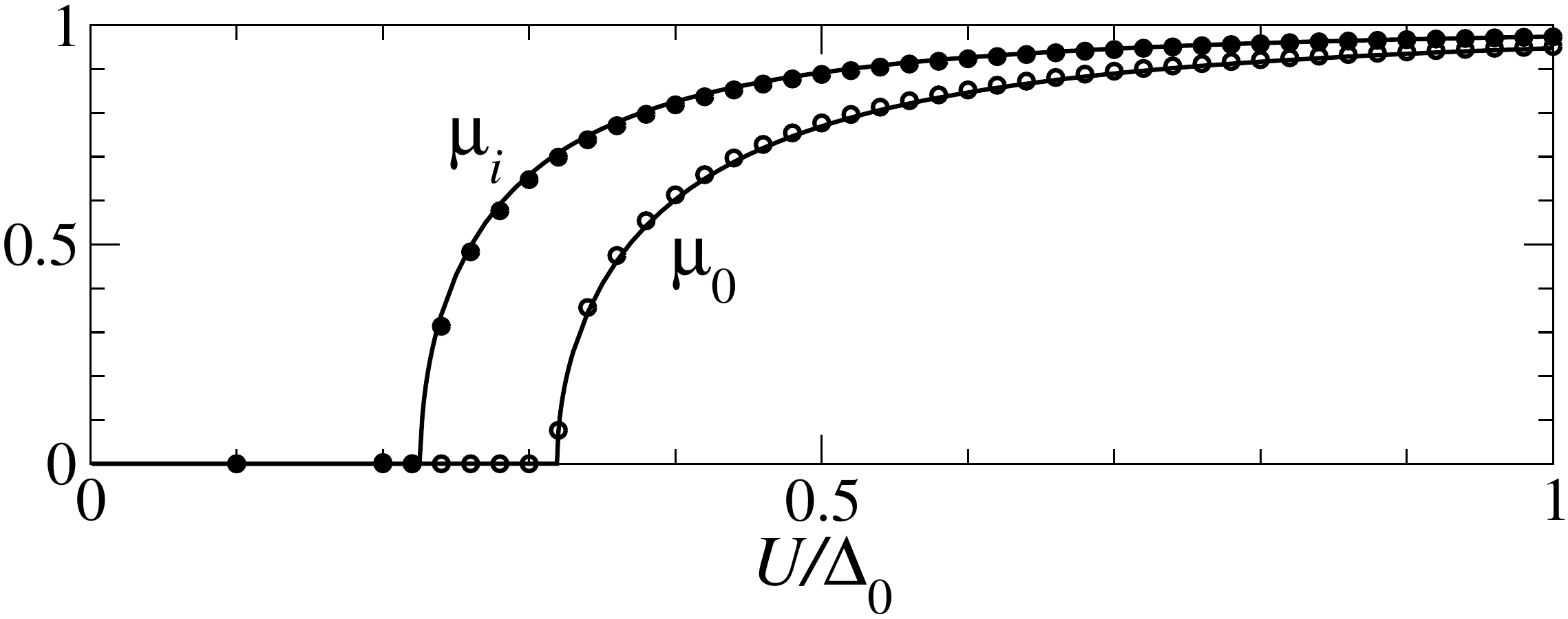}

  \caption{Magnetic impurity moment vs.~Coulomb repulsion $U$ for an
    AIM with bandwidth $D=0.01\Do$, where $\wo\simeq 0.08\Do$.  Filled
    circles: numerically determined LMA moment $\mi$, as required by
    eq.~(\ref{pinning}); open circles: numerically determined UHF
    moment $\mo$. Solid lines: corresponding approximate analytical
    expressions, eqs.~(\ref{mulma_fit}) and (\ref{muuhf_twosite}). The
    critical interactions are $\Uc\simeq 0.23\Do$ for the LMA, and
    $\Uco\simeq 0.32\Do$ in UHF.}
  \label{fig:uvsmu} 
  \end{center}
\end{figure}
In Fig.~\ref{fig:uvsmu}, the self-consistent UHF moment $\mo$,
eq.~(\ref{scmu}), is plotted as a function of the interaction strength
$U$ for a narrow-band AIM with host bandwidth $D=0.01\Do$.

The spectral density $\DAuhf{\dn}(\w)$ consists of a low-energy
continuum for $\abs{\w}<D$, of net weight $\order{D/\Do}$, arising
from the finite imaginary part of the hybridization, and two pole
contributions, one above the single-particle band, at $\w=\w_{>}>+D$,
and the other one below it, at $\w=-\w_{<}<-D$.  Assuming both poles
to occur far outside the single-particle band, the pole frequencies
and weights can be obtained rather accurately by expansion of the
hybridization function (\ref{nbhybrid}), $\ReDelta(\w)\sim(2/\pi)\Do
D/\w$:
\begin{subequations}
\label{pole_gl}
\begin{eqnarray}
\label{omega_gl}
\w_{\gtrless} & \simeq & \wo\,\left( \sqrt{y_{0}^2+1}\pm y_{0} \right)
\\
\label{weight_gl}
q_{\gtrless} & \simeq & \frac{ \sqrt{y_{0}^2+1}\pm y_{0}}{2 \sqrt{y_{0}^2+1}}
\end{eqnarray}
\end{subequations}
with $y_{0}=U\mo/4\wo$.  

For the above expansion to hold, ${\w_{\gtrless}}\gg D$ is required,
thus limiting the validity of eq.~(\ref{pole_gl}) for the low-energy
pole at $\w=-\w_{<}$ to interactions $U\mo\ll\frac{4}{\pi}\Do$ (while
no restrictions follow for the high-energy pole $\w_{>}$). In this
range of interactions, the renormalization effects in the
single-particle continuum are still small and the UHF moment $\mo$ can
be obtained very accurately by only retaining the pole contributions
of $\DAsuhf(\w)$ in the self-consistency equation (\ref{scmu}). This
yields the following expression which corresponds to the lower solid
line in Fig.~\ref{fig:uvsmu}:
\begin{equation}
\label{muuhf_twosite}
\mo\,\simeq\,
\begin{cases}
\sqrt{1-\left({\Uco}/{U}\right)^{2}} & \mbox{for $U>\Uco:=4\wo$} \cr
0 & \mbox{for $U<\Uco$}
\end{cases}
\end{equation}

For $U<\Uco$, the self-consistently determined moment vanishes and the
UHF Green function coincides with the noninteracting or RHF solution,
eq.~(\ref{Grhf}).  Above $\Uco$, by contrast, a finite local moment
forms on the impurity, and saturates rapidly as $U$ is increased.
With increasing $\mo$, the pole at $\w=\w_{>}$ shifts rapidly away
from its noninteracting value $\wo$ towards higher frequencies and
gains in intensity. For $U\gg\Uco$, it becomes the upper Hubbard
satellite at $\w_{>}\simeq U/2$, which overwhelmingly dominates the
spectrum $\DAuhf{\dn}(\w)$ with a pole weight of $q_{>}\simeq
1-\order{[\wo/U]^{2}}$. Simultaneously, the pole at $\w=-\w_{<}$ moves
from $-\wo$ towards the lower band edge $-D$ and loses weight.  For
$\Uco\ll U \lesssim \Do$, its position and weight are given to good
accuracy by
\begin{subequations}
\label{poles_l_medU}
\begin{eqnarray}
\label{omega_l_medU}
\w_{<}&\simeq&\frac{2\wo^{2}}{U}\,=:\, \frac{J}{2} \;\mbox{,}\\
\label{weight_l_medU}
q_{<}&\simeq& \left(\frac{2\wo}{U}\right)^{2}
\;\mbox{.}
\end{eqnarray}
\end{subequations}
The first equation defines a third low-energy scale, $J=4\wo^{2}/U$,
which lies between bandwidth and bonding energy, $D\ll J\ll \wo$,
and accounts for the antiferromagnetic exchange between impurity and
host.

If $U$ is increased to values of the order of $\Do$, the
antiferromagnetic exchange $J$ approaches the bandwidth $D$, and the
above analysis --- while still valid for the Hubbard level at
$\w=\w_{>}$ --- breaks down for the low-energy pole at $\w=-\w_{<}$.
If, like in the present case, the hybridization band $\ImDelta(\w)$
has a discontinuity at the lower band edge $-D$, the logarithmic
divergence of the related real part, $\ReDelta(\w)$, still guarantees
the existence of a low-energy pole; in the present model with a flat
hybridization band it occurs, for $U\gg\Do$ (where $\mo\simeq 1$),
exponentially close to the band edge and carries exponentially small
weight, which renders it insignificant in practice:
\begin{subequations}
\label{poles_l_lrgU}
\begin{eqnarray}
\label{omega_l_lrgU}
\w_{<} &\simeq& 
D\left[ 1\,+\,2 \exp\left(-\frac{\pi U}{2\Do}\right)\right]
\\
\label{weight_l_lrgU}
q_{<} &\simeq& \frac{2\pi D}{\Do} \exp\left(-\frac{\pi U}{2\Do}\right) 
\end{eqnarray}
\end{subequations}

\begin{figure}[htbp]
  \begin{center} 
  \leavevmode 
  \includegraphics[width=\FigureWidth]{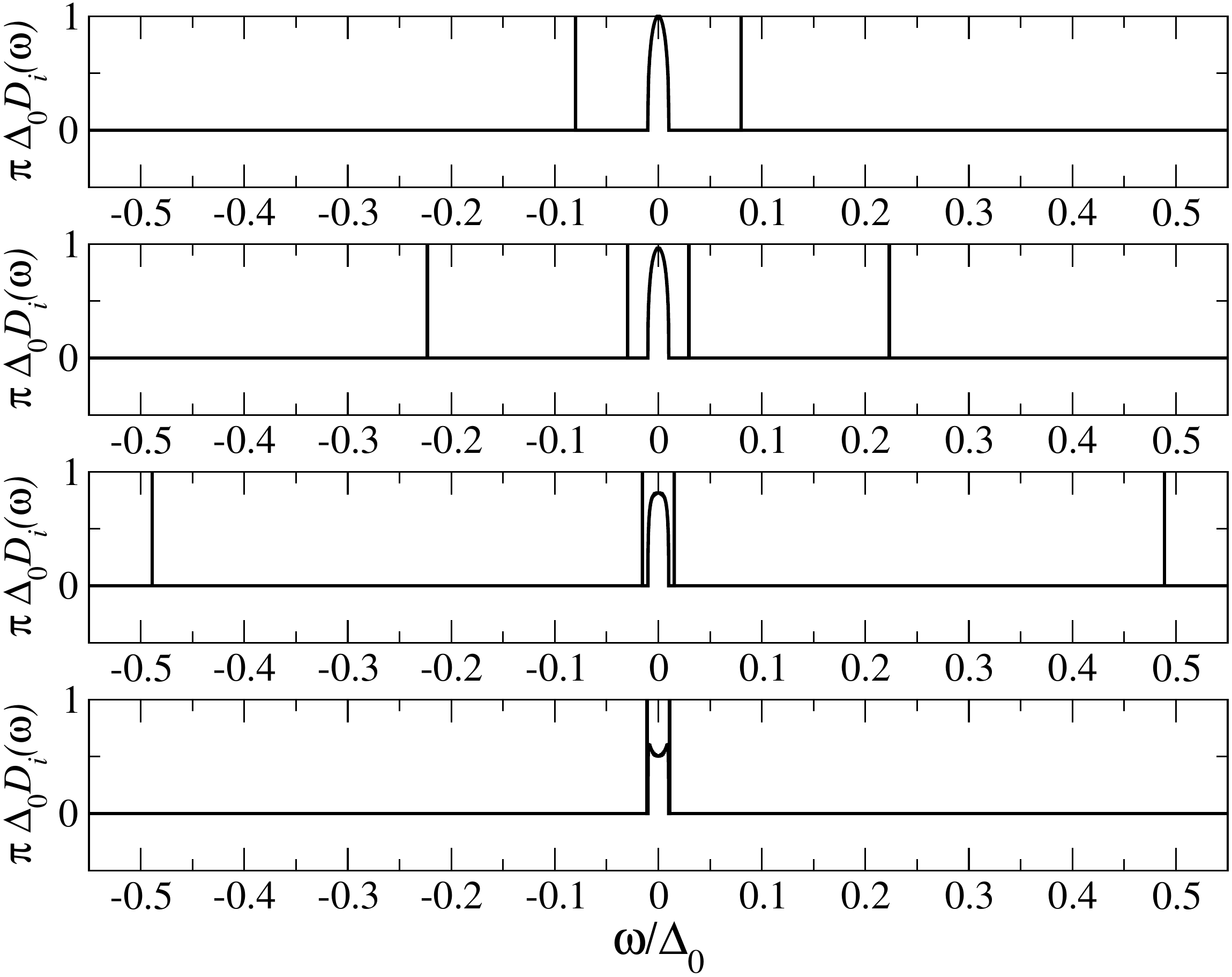}

  \caption{UHF impurity spectra, $\pi\Do\DiUHF(\w)$ vs.~$\w/\Do$, for
    bandwidth $D=0.01\Do$, {\sl i.e.}, $\wo\simeq 0.08\Do$ and
    $\Uco\simeq 0.32\Do$, and interaction strengths $U/\Do=0$, $0.5$,
    $1$, $2$ (from top to bottom).  In the bottom panel, the Hubbard
    satellites at $\w\simeq \pm U/2=\pm\Do$ lie off plot
    range. Discrete levels are represented by vertical lines. The
    Fermi energy is $\w=0$.}

  \label{fig:dosuhf} 
  \end{center}
\end{figure}

Finally, according to eq.~(\ref{Guhfmix}), the full UHF impurity
spectrum can be obtained by superposing $\DAuhf{\dn}(\w)$ and
$\DBuhf{\dn}(\w)$, the latter of which follows by symmetry,
$\DBuhf{\dn}(\w)=\DAuhf{\up}(\w)$. As illustrated in
Fig.~\ref{fig:dosuhf}, $\DiUHF(\w)$ consist of a continuum for
$\abs{\w}<D$, and two pairs of poles at $\w=\pm\w_{>}$ and
$\w=\pm\w_{<}$.  Before the onset of moment formation, {\sl i.e.},  for
$U<\Uco$, the two pairs merge into the single pair of molecular
orbitals shown in the first graph, at $\w=\pm\wo$ and of net weight
$q\sim 1/2$ each, and UHF coincides with the noninteracting solution.
As $U$ exceeds $\Uco$, the orbitals split up progressively into a
stronger growing high- and a weaker growing low-energy component, as
depicted in the second and third graphs of Fig.~\ref{fig:dosuhf}. In the second
panel, where $U=0.5\Do$ is only moderately greater than $\Uco\simeq
0.32\Do$, the impurity moment is already well established, $\mo\simeq
0.78$, and the ``Hubbard satellites'' at $\w\simeq\pm 0.22\Do$ are not
far from their terminal position, $\w\simeq\pm U/2=\pm 0.25\Do$,
carrying together more than $88\%$ of the spectral
intensity. Approximately a further $11\%$ of the weight reside in the
low-frequency poles at $\w\simeq\pm 0.030\Do$ (expected at
$\w\simeq\pm J/2\simeq\pm 0.025\Do$ from eq.~(\ref{omega_l_medU})),
while the central low-energy continuum carries only less than $1\%$ of
the spectral intensity.
In the third graph, where $U=\Do$, roughly $98\%$ of the spectral weight
reside in the Hubbard satellites at $\w\simeq\pm 0.49\Do$, while the
low-energy poles at $\w\simeq\pm 0.015\Do$ appear close to the central
continuum, the latter being strongly renormalized from its
noninteracting shape.
Finally, these renormalization effects are still enhanced for $U=2\Do$
(bottom panel), not yet fully in the strong coupling regime, resulting
in a considerable violation of the Friedel sum rule (see below); here,
almost all intensity resides in the Hubbard satellites (off plot
range), and exponentially weak low-energy poles, with total weight
$0.3\%$, are located slightly outside the central continuum [see
eqs.~(\ref{poles_l_lrgU})].

The above scenario, with a spectral evolution governed by
$\wo\simeq\sqrt{2\Do D/\pi}$ in weak coupling, and by $U$ and
$J=4\wo^2/U$ in moderate coupling, $\Uco\ll U\lesssim \Do$, concurs
qualitatively with second order perturbation theory in $U$ (2PT),
\cite{SL01} and can be rationalized in terms of a simple two-site
model in which the host is caricatured by a single level coupled to
the impurity. \cite{Lange98} Despite these encouraging results, UHF
suffers from several severe limitations: (i) in the two-site limit,
$D\to 0$ albeit with finite $\wo$, exactly captured by 2PT, the
low-energy single-particle levels are expected at $\w\simeq \pm 3J/2$
for $U\gg\Uco$, {\sl i.e.}, three times the corresponding UHF pole
frequency (\ref{omega_l_medU}); (ii) the UHF single-particle band,
present for $\abs{\w}<D$, is generally not a Fermi liquid: for
interactions above $\Uco$, a nonzero impurity moment $\mo$ forms, and
the UHF zero-frequency behaviour
$\DiUHF(\w=0)=(1/\pi\Do)/[1+(U\mo/2\Do)^{2}]$ violates the Friedel
sum rule\cite{Hewson,Langreth66} that compels the impurity spectra of
the particle-hole symmetric AIM, for arbitrary interaction strengths,
to be pinned at the Fermi level to their noninteracting value,
$\Di{}(\w=0)=1/\pi\Do$; and (iii), relatedly, due to the absence of
dynamics UHF completely fails to capture any of the Kondo physics
expected to govern the low-energy spectrum in the strong coupling
regime, $U\gg 4\Do$.

The present paper aims to transcend these shortcomings within the
framework of the local moment approach (LMA).

% %%%%%%%%%%%%%%%%%%%%%%%%%%%%%%%%%%%%%%%%%%%%%%%%%%%%%%%%%%%%
\section{Local Moment Approach (LMA)}
\label{sec:lma}

The LMA\cite{Logan98,Glossop02,Dickens01} expresses the impurity Green
function in a formalism employing two self-energies. At pure
mean-field level this description reduces to UHF, discussed in the
previous section, with each of the self-energies arising from
impurities with predominant $\up$- or $\dn$-spin occupation. For an
impurity hosted by an insulator with a sufficiently large gap, this
doubly degenerate ground-state is actually
observed. \cite{Takegahara92,Chen98,Galpin08a,Galpin08b} Conversely,
for a metallic host, dynamical tunneling between the two
broken-symmetry states can (and will) still lower the energy and
ultimately lead to the formation of a ``Kondo'' singlet ground-state
with fully restored spin symmetry on the longest
timescales.\cite{Logan98,Glossop02} The tunneling mechanism requires a
(virtually) doubly occupied or empty impurity at some intermediate
stage, implying that its rate -- and the corresponding energy scale --
diminish with increasing interaction strength. The LMA incorporates
such a mechanism by coupling the single-particle dynamics to
energetically low-lying flips of the impurity moment. In order to
obtain a successful description of insulating and metallic phases
within the same framework, this has to be done in a manner
encompassing the possibility of self-consistent restoration of the
spin symmetry at low energies, as necessary for the preservation of
Fermi-liquid behaviour on this scale.

The implementation of the LMA follows Refs.~\onlinecite{Logan98} and
\onlinecite{Glossop02}. For reasons analogous to those discussed in
the context of eq.~(\ref{Guhfmix}), the full impurity Green function
$\Gi{}$ is again obtained by superposing ${\rm A}$- and ${\rm B}$-type
impurity propagators in a spin-rotationally invariant fashion,
\begin{subequations}
  \label{Gmix}
  \begin{equation}
    \label{Gmixtype}
    \Gi{}(\w)\,=\,
    \frac{1}{2}\,\left[\GA{\sigma}(\w)\,+\,\GB{\sigma}(\w)\right]
    \;\mbox{.}
  \end{equation}
  Making use of
  spin symmetry, $\GB{\sigma}(\w)=\GA{-\sigma}(\w)$, yields the
  equivalent expression
  \begin{equation}
    \label{Gmixspin}
    \Gi{}(\w)\,=\,
    \frac{1}{2}\,\left[\Gal{\up}(\w)\,+\,\Gal{\dn}(\w)\right]
    \;\mbox{.}
  \end{equation}
\end{subequations}
Here, $\Gi{}$ is independent of the impurity type and the indices
$\alpha$ may be suppressed for convenience, meaning that the
individual (broken-symmetry) $\Gas(\w)$ in eq.~(\ref{Gmixspin}) will
henceforth be implicitly considered ``${\rm A}$-type'', {\sl i.e.}, of
impurity moment $\mi\ge 0$, unless stated otherwise. Each of the
$\Gas(\w)$ can be expressed in terms of a Dyson equation,
\footnote{In the present particle-hole symmetric AIM, where the
  impurity level is chained to the interaction strength {\sl via}
  $\epsi=-U/2$, it is convenient to include $\epsi$ in the
  self-energies $\fullSa{\sigma}(\w)$ instead of the noninteracting
  $\GiF(\w)$. For the general (asymmetric) AIM, where $\epsi$ and $U$
  are truly independent, the opposite choice, used in
  Ref.~\onlinecite{Glossop02}, is appropriate.}
\begin{equation}
  \label{Galpha1}
  \Ga{\sigma}(\w)\,=\,\left[\GiF^{-1}(\w)\,-\,\fullSa{\sigma}(\w)\right]^{-1}
  \;\mbox{,}
\end{equation}
thus defining the two self-energies $\fullSa{\up}(\w)$ and
$\fullSa{\dn}(\w)$ central to the present approach.  Without loss of
generality, the two self-energies can be separated into static and
dynamic contributions, $\fullSa{\sigma}(\w)=\Sa{\sigma}^{\rm
  st}+\Sa{\sigma}(\w)$: diagrammatically, the former are suitably
approximated by the Hartree tadpole (while the contribution of the
Fock open oyster diagram vanishes), amounting to
\begin{equation}
  \label{SigmaStatic}
  \Sa{\sigma}^{\rm st}\,=\,
  -\sigma\frac{U}{2}\int\limits_{-\infty}^{0} {\rm d}\w 
  \left[\DAuhf{\up}(\w)-\DAuhf{\dn}(\w)\right]\;\mbox{;}
\end{equation}
and the dynamical $\Sa{\sigma}(\w)$ are defined to contain
``everything else''.  Eq.~(\ref{Galpha1}) may thus be rephrased as
\begin{equation}
  \label{Galpha}
  \Ga{\sigma}(\w)\,=\,
  \left[\Gauhf{\sigma}^{-1}(\w)\,-\,
    \left(\Sa{\sigma}^{\rm st}-\epsis\right)\,-\,
    \Sa{\sigma}(\w)\right]^{-1}
  \;\mbox{.}
\end{equation}  
This choice is particularly convenient if -- as for the LMA -- the
dynamic self-energy contributions $\Sa{\sigma}(\w)$ are to be
diagrammatically constructed from the UHF propagators $\Gasuhf$ rather
than from the noninteracting $\GiF$, since it entails the sum of the
static contributions (in braces) to vanish {\em if} the UHF moment
$\mo$, determined self-consistently from eq.~(\ref{scmu}), is used.

%But even for small departures of the impurity moment $\mi$ from its
%UHF value -- as will usually be required for symmetry restoration (see
%below) -- the contribution in braces remains minute and can be safely
%neglected in the following.

The present two-self-energy description is by now well established and
by no means exclusive to the LMA, but likewise emerges in other
approaches as, e.g., in the NRG where it occurs for odd iterations of
the renormalization group.\cite{Galpin05} Nevertheless, it is
desirable to connect it to the conventional single self-energy,
defined by the Dyson equation
$\Gi{}(\w)=[\GiF^{-1}(\w)-\Si{}(\w)]^{-1}$:\cite{Logan98,Glossop02}
\begin{eqnarray} 
  \label{singleSigma}
  \Si{}(\w) &=&
  \sfrac{1}{2}\left[\fullSa{\up}(\w)+\fullSa{\dn}(\w)\right]
  \nonumber
  \\
  && +
  \frac{\left(
      \frac{1}{2}\left[\fullSa{\up}(\w)-\fullSa{\dn}(\w)\right]
    \right)^{2}}{
    \GiF^{-1}(\w)-\frac{1}{2}\left[\fullSa{\up}(\w)+\fullSa{\dn}(\w)\right]
  }
\end{eqnarray}

Before specifying the class of diagrams to be retained to approximate
the dynamical self-energies $\Sa{\sigma}(\w)$, the conditions
necessary for Fermi-liquid behaviour to prevail at low frequencies
shall be reviewed.\cite{Glossop02} According to
Luttinger,\cite{Luttinger61} the imaginary part of the {\em single}
self-energy $\Si{}(\w)$ is required to vanish as $\order{\w^2}$ at the
Fermi level $\w=0$, which for the two self-energies
$\fullSa{\sigma}(\w)=\RefullSa{\sigma}(\w)-\ic\,\sign{\w}\ImfullSa{\sigma}(\w)$
employed by the present approach translates to the following two
conditions:
\begin{subequations}
  \label{LuttingerCond}
  \begin{eqnarray}
    \label{LuttingerIm}
    && \ImfullSa{\sigma}(\w)\;\sim\;\order{\w^{2}}\quad\mbox{for $\w\to0$;}
    \\
    \label{LuttingerRe}
    && \RefullSa{\up}(\w=0)\;=\;\RefullSa{\dn}(\w=0)\;\mbox{.}
  \end{eqnarray}
\end{subequations}
The fulfilment of the first condition shall be assumed from now on,
and will be explicitly shown in Sec.~\ref{sec:selfenergy} for the class
of diagrams chosen in the following. The second condition requires the
broken symmetry $\fullSa{\sigma}(\w)$ to coincide with the fully
symmetric single $\Si{}(\w)$ at the Fermi level, thus expressing the
concept of a restored spin symmetry on the lowest energy
scales\cite{Glossop02} which was alluded to in the beginning of this
section.  Here, the additional particle-hole symmetry,
$\fullSa{\up}(\w)=-\fullSa{\dn}(-\w)$, in combination with
eqs.~(\ref{LuttingerCond}), require both self-energies
$\fullSa{\sigma}(\w)$ to vanish at the Fermi level. This, in turn,
{\em automatically} guarantees the Friedel sum-rule pinning of the
spectra, $\Di{}(\w=0)=1/\pi\Do$.
%(as opposed to the general asymmetric AIM, where the Friedel sum rule
%has to be implemented separately).

In its self-energy diagrams, the LMA incorporates transverse spin
fluctuations responsible for a dynamical reversion of the impurity
moment. At the simplest level, to be implemented in the following,
this is achieved by coupling the ladder-sum polarization propagator to
the single-particle Green function,
\begin{equation}
  \label{fig:SigmaLMA}
  \leavevmode
  \unitlength=2ex 
  \setlength{\Units}{\the\unitlength} 
  \begin{picture}(12,6)
    \put( 0,3){\makebox(0,0)[c]{$\Sa{\sigma}(\w)\equiv\Sa{\sigma}[\{\Gasuhf\}]\,\sim\,$}}
    \put( 6,1){\includegraphics[width=5\Units]{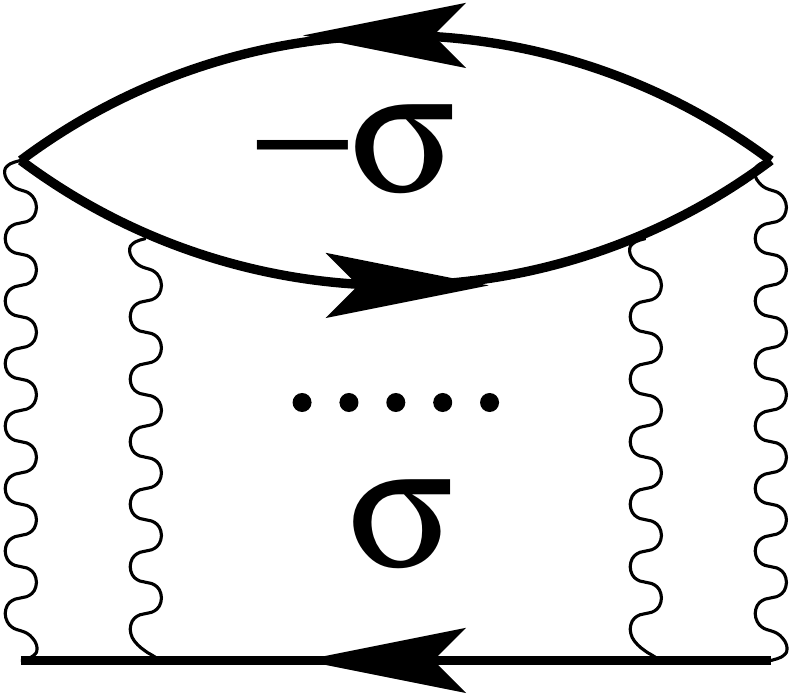}}
    \put(12,3){\makebox(0,0)[c]{$\equiv$}}
    \put(13,0){\includegraphics[width=5\Units]{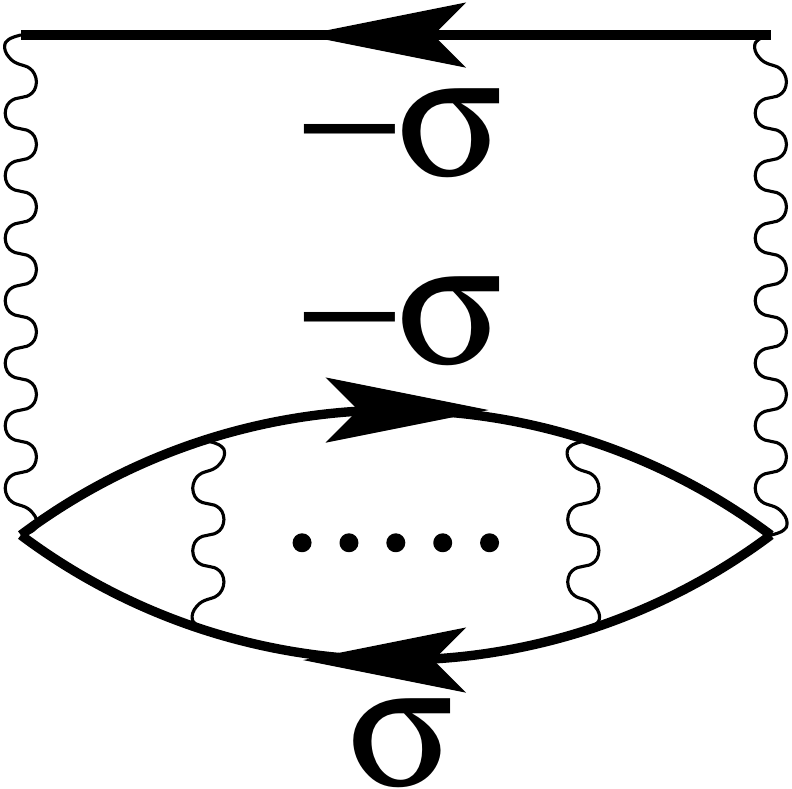}}
  \end{picture}
\end{equation}
where wavy lines represent the interaction $U$ and left-going
(right-going) solid lines stand for a UHF-like particle (hole)
propagator $\Gasuhf$. Note, however, that symmetry restoration is
generally not automatically guaranteed by the above class of diagrams,
but can be obtained through self-consistent determination of a free
parameter. In the following, the impurity moment $\mi$ will serve as
such, implying that it will generally differ at self-consistency from
its UHF value $\mo$ (see Sec.~\ref{sec:selfenergy} below).  

The above approximation for the self-energies $\Sa{\sigma}(\w)$ has
been motivated and discussed in detail in
Ref.~\onlinecite{Logan98,Glossop00,Glossop02}; here, its basic
properties shall only be briefly reviewed:
(i) in weak coupling, before the onset of moment formation, the UHF
propagators $\Gauhf{\sigma}$ coincide, for both $\sigma$, with the
noninteracting Green functions $\GiF$; up to (and including) second
order in $U$, eq.~(\ref{fig:SigmaLMA}) is hence equivalent to an
ordinary diagrammatic perturbation expansion about the noninteracting
limit.
(ii) For arbitrary interaction strengths, the self-energy diagram
(\ref{fig:SigmaLMA}) describes processes where after having added at
$t=0$, say, a $\sigma=\dn$-spin electron to a $-\sigma=\up$-spin
occupied impurity, the original $\up$-spin electron hops away, thus
effectively flipping the impurity spin until its return at some later
time $t$: this is precisely the dynamical spin-flip scattering,
comprised in the second order of Anderson's {\em poor man's scaling}
(see Ref.~\onlinecite{Hewson}), which allows us to restore spin symmetry
and to capture the strong coupling Kondo limit.
(iii) In the atomic limit, where $V_{i\vk}\equiv 0$ and therefore
$\Do=0$, the self-energy (\ref{fig:SigmaLMA}) vanishes, since the
$-\sigma$-spin electron cannot leave the impurity at all. The LMA
reduces to simple UHF, and thus becomes exact in this limit,
predicting an impurity moment of $\mo=1$ and single-particle poles at
$\w=\pm U/2$.  This is salutary, since the atomic limit may be
considered as an extreme of the local moment phases observed in a
gapped AIM; \cite{Takegahara92,Chen98,Galpin08a,Galpin08b} it also
plays an important role in the context of the breakdown of the
skeleton expansion\cite{SL01} taking place already in the metallic
regime close to a Mott metal-insulator
transition. \cite{Kehrein98,Hofstetter99}

Before discussing the LMA impurity spectra in Sec.~\ref{sec:spectra}, the
spin fluctuations and their dynamical coupling to the single-particle
Green function will be examined in the following two paragraphs.

% %%%%%%%%%%%%%%%%%%%%%%%%%%%%%%%%%%%%%%%%
\subsection{Transverse spin fluctuations}
\label{sec:Pi}

As already mentioned in the previous paragraph, one of the main
ingredients for the LMA self-energy is given by dynamical spin-flip
scattering processes opening the possibility to restore the spin
symmetry on the lowest energy scale. Embodied in the transverse
impurity-spin polarization propagators,
$\Pipm{}(t)=\ic\mean{\hat{T}\{S_{i}^{+}(t)S_{i}^{-}(0)\}}$ and
$\Pimp{}(t)$, these scattering processes are at the simplest level
accounted for by a sum of repeated particle-hole interactions, as
depicted by the ladder bubble in the rightmost self-energy diagram in
eq.~(\ref{fig:SigmaLMA}) which, for $\sigma=\dn$ ($-\sigma=\up$),
leads to the familiar expression
\begin{equation}
\label{PiRPA}
\Piapm(\w)\,=\,\frac{\Poapm(\w)}{1-U\Poapm(\w)}
\;\mbox{,}
\end{equation}
where 
\begin{equation}
\label{Pi0}
\Poapm(\w) \,=\, 
\ic \int\limits_{-\infty}^{+\infty} \frac{{\rm d}\w'}{2\pi}\, 
\Gauhf{\dn}(\w')\,\Gauhf{\up}(\w'-\w)
\end{equation}
is the bare transverse-spin bubble. [The second polarization propagator
follows by particle-hole symmetry,  $\Piapm(\w)=\Piamp(-\w)$.]

Using the Hilbert transform relation (\ref{Hilbert}) for $\Gasuhf(\w)$
on the right-hand side (rhs) of eq.~(\ref{Pi0}), yields \cite{Logan98}
\begin{eqnarray}
\label{ImPi0}
\frac{1}{\pi} \imag\Poapm(\w) \,=\,   
\theta(\w) && \int\limits_{0}^{+\abs{\w}} {\rm d}\w'\,
\Dauhf{\dn}(\w')\,\Dauhf{\up}(\w'-\w)
\nonumber \\
+\theta(-\w) && \int\limits_{-\abs{\w}}^{0} {\rm d}\w'\,
\Dauhf{\dn}(\w')\,\Dauhf{\up}(\w'-\w)
\mbox{.}
\nonumber \\
&&
\end{eqnarray}
The first term describes a spin-flipping particle-hole excitation
where an electron is removed from an occupied $\up$-spin state below
the Fermi surface ($\w'-\w<0$) and placed into an unoccupied
$\dn$-spin state above the Fermi surface ($\w'>0$).  Similarly, the
second term accounts for the recombination of an $\up$-spin electron
into an empty $\dn$-spin level below the Fermi surface. 

In the present AIM with a narrow host band, these processes operate
mainly between the UHF orbitals, giving rise to poles in
$\frac{1}{\pi}\imag\Poapm(\w)$ at $\w=2\w_{>}$ and $\w=-2\w_{<}$ with
respective pole weight $q_{\gtrless}^{2}$.

For a vanishing impurity moment, as appropriate for weak coupling, the
UHF levels coincide with the ``molecular'' orbitals of the
noninteracting limit, entailing poles in
$\frac{1}{\pi}\imag\Poapm(\w)$ at frequencies $\w=\pm 2\wo$ and of
spectral weight $q\simeq 1/4$ each.
Conversely, in moderate to strong coupling, particle-hole excitations
between the lower and the upper UHF Hubbard level become dominant,
producing a high-frequency pole at $\w=2\w_{>}\simeq U$ of weight
$q_{>}^{2}\simeq 1$ in $\frac{1}{\pi}\imag\Poapm(\w)$, and a
corresponding low-frequency pole at $\w=-2\w_{<}$ whose intensity
decreases rapidly (first algebraically for $\Uc\ll U\lesssim \Do$,
where $q_{<}^{2}\sim [2\wo/U]^{4}$, and then exponentially for $U\gg
4\Do$).

In addition to the pole contributions $\frac{1}{\pi}\imag\Poapm(\w)$
contains three weak continua stemming from processes involving the
narrow UHF single-particle band at the Fermi level.
Two of these continua have their origin in electronic transfer between
the single-particle band and one of the UHF levels; they are found at
$\w\in [\w_{>},\w_{>}+D]$ and at $\w\in [-\w_{<}-D,-\w_{<}]$, with
respective net weight $\order{q_{\gtrless} D/\Do}$, and turn out to be
of little importance in the following.
The third continuum appears at low frequencies, $\abs{\w}\le 2D$, and
owes its existence to particle-hole excitations within the narrow UHF
single-particle band itself.  According to eq.~(\ref{ImPi0}), its
behaviour in the vicinity of the Fermi level is essentially governed
by the overlap of the two $\Dasuhf(\w)$-bands; for a metallic host,
the intensity of the latter being finite at the Fermi level $\w=0$,
this results in a continuum vanishing linearly for small $\w$,
\begin{equation}
\label{ImPi0_omega0}
\frac{1}{\pi}\imag\Poapm(\w) \, \stackrel{\abs{\w}\ll D}{\sim} \,
\abs{\w} \Dauhf{\dn}(0)\Dauhf{\up}(0)
\;\mbox{,}
\end{equation}
a property that is needed to prove the fulfilment of condition
(\ref{LuttingerIm}) by the present theory. \cite{Logan98}

\subsubsection{Stability of the ladder sum propagator.}
\label{sec:RPAstability}

From the Hilbert transform relation (\ref{Hilbert}) for $\Piapm(\w)$,
and the positivity of its imaginary part, one deduces an analyticity
condition for the polarization propagator, \cite{Logan98}
\begin{equation}
\label{RePi_omega0}
\real\Piapm(\w=0)\,=\,
\int\limits_{-\infty}^{+\infty} \frac{{\rm d}\w'}{\pi}\,
\frac{\imag\Piapm(\w')}{\abs{\w'}}\,>\,0
\;\mbox{,}
\end{equation}
which naturally holds for $\Poapm$ and the exact $\Piapm$.

Conversely, for an approximate polarization propagator, like the
present ladder-sum $\Piapm$, eq.~(\ref{RePi_omega0}) is not
automatically satisfied, but rather constrains the applicability of the
approximation to a certain range of parameters $U$ and
$\mi$. 

According to eq.~(\ref{PiRPA}) and granted $\imag\Poapm(\w=0)=0$, the
behaviour of $\real\Piapm(\w=0)$ is solely controlled by the bare
$\real\Poapm(\w=0)$.  For a vanishing impurity moment, as enforced in
RHF and found self-consistently in UHF for $U<\Uco\equiv 4\wo$,
$\real\Poapm(\w=0)\simeq 1/4\wo$ follows to good accuracy from the
two-site approximation, implying that the corresponding ladder-sum
$\Piapm$ satisfies the positivity condition (\ref{RePi_omega0}).

Above $\Uco$, the $\mi=0$ ladder-sum $\Piapm$ --- which corresponds to
an ordinary random phase approximation (RPA) in the transverse spin
channel --- becomes unstable to spin excitations since the underlying
mean-field theory predicts a transition to a phase with a finite
magnetic moment.  A finite $\mi$, on the other hand, allows for a
direct calculation of $\real\Poapm(\w=0)$ from eq.~(\ref{Pi0}) by
means of the identity
$\Gauhf{\up}(\w)-\Gauhf{\dn}(\w)=-U\mi\Gauhf{\dn}(\w)\Gauhf{\up}(\w)$
and of particle-hole symmetry
$\Dauhf{\sigma}(\w)=\Dauhf{-\sigma}(-\w)$:
\begin{equation}
\label{RePi0_omega0}
U \real\Poapm(\w=0)\,=\,
\frac{1}{\mi}\,\int\limits_{-\infty}^{0} {\rm d}\w 
\left[ \Dauhf{\up}(\w)- \Dauhf{\dn}(\w)\right]
\;\mbox{.}
\end{equation}
The latter equation shows that if the moment is determined
self-consistently from the UHF propagators, $\mi=\mo$ as given in UHF
by eq.~(\ref{scmu}), the rhs evaluates to unity and the corresponding
ladder-sum $\Piapm$, eq.~(\ref{PiRPA}), has a pole at $\w=0$.  While
appropriate for an insulator, where flipping the impurity spin costs
no energy, in a metallic host such a spin flip is expected to be
governed by the Kondo effect and thus to involve a finite energy.

In the framework of the LMA, this picture emerges naturally if $\mi$
is increased above its UHF value $\mo$, thereby shifting the pole in
$\Piapm(\w)$ from $\w=0$ to a small but positive frequency $\wm$ ---
closely related to the Kondo energy --- and rendering $\Piapm$
analytic in the sense of eq.~(\ref{RePi_omega0}).

\subsubsection{Sum rules and renormalized ladder-sum propagator.}
\label{sec:sumrule}

Any acceptable approximation for the spin-flip polarization propagator
should fulfil the following sum rules:
\begin{eqnarray}
\label{Pisumrule}
\left[ \int\limits_{0}^{+\infty} \!\!\!\pm\!\!\! \int\limits_{-\infty}^{0} \right]
\frac{{\rm d}\w}{\pi} \imag\Piapm(\w)
& = & \mean{\comm{c^{+}_{i\up} c_{i\dn}}{c^{+}_{i\dn} c_{i\up}}_{\pm}}
\nonumber \\
& = &
\begin{cases} 
n_{i\up}+n_{i\dn}-2\mean{\hat{n}_{i\up}\hat{n}_{i\dn}} & \cr
       \mi & \cr
\end{cases}
\nonumber \\
\end{eqnarray}
which naturally hold for the noninteracting $\Poapm(\w)$, and
likewise for the exact $\Piapm(\w)$.

The second sum rule reflects the fact that, in weak coupling,
$\imag\Piapm(\w)$ is symmetric about the Fermi level since the
impurity is found to be occupied by electrons of both spin species
with equal probability; after the onset of moment formation (and
before symmetry restoration), the impurity is predominantly $\up$-spin
occupied and spectral intensity at positive energies, associated with
processes flipping the impurity spin from $\up$ to $\dn$, should
dominate.

As for the first sum rule, its rhs is comprised between $1/2$ and $1$
for the particle-hole symmetric AIM, since the expectation to find a
doubly occupied impurity $\mean{\hat{n}_{i\up}\hat{n}_{i\dn}}$ varies
from $1/4$ in weak to $0$ in strong coupling and, in any case,
$n_{i\up}+n_{i\dn}=1$; both limits are captured by the approximate
expression $\frac{1}{2}[1+\mi^2]$ which follows by recasting the
double occupancy as
$\mean{\hat{n}_{i\up}\hat{n}_{i\dn}}=\frac{1}{4}(\mean{\hat{n}_{i}^{2}}-\mean{\hat{\mu}_{i}^{2}})$
and subsequently factorizing the quartic Fermion operators via a
Hartree-Fock decoupling.

For the present narrow-band AIM, the above sum rules are naturally
fulfilled in weak coupling, where $\Piapm(\w)$ reduces essentially to
the noninteracting $\Poapm(\w)$. More surprisingly, also in the
strong coupling (Kondo) regime, $U\gg 4\Do$, both sum rules are found
to be approximately satisfied for values of the magnetic moment
required by symmetry restoration.

Problems are found to arise primarily for moderate interactions, large
enough to find the impurity moment well established, but far from the
Kondo limit.  For this range of interactions, the main features of the
ladder-sum polarization propagator $\Piapm(\w)$ are accurately
described in the two-site approximation; it reduces to two poles, with
frequencies and weights given analytically by
\begin{subequations}
  \label{pole_rpa}
  \begin{eqnarray}
    \label{omega_rpa}
    \wm^{\pm}& = & 2\wo\left[\gamma y \pm \sqrt{(\gamma y)^{2}+\gamma}\right]
    \;\mbox{,}
    \\
    \label{weight_rpa}
    \qm^{\pm}& = & 
    \frac{\left[\gamma y \pm \sqrt{(\gamma y)^{2}+\gamma}\right]^{2}}
    {4\gamma\sqrt{\left[y^{2}+1\right]\left[(\gamma y)^{2}+\gamma\right]}}
    \;\mbox{,}
  \end{eqnarray}
\end{subequations}
where $y=\mu U/\Uco$ and $\gamma=1-\frac{U/\Uco}{\sqrt{y^{2}+1}}$.
Incidentally, eqs.~(\ref{pole_rpa}) encompass the bare (broken
symmetry) $\Poapm(\w)$, which follows for $\gamma=1$, as well as the
noninteracting polarization bubble, obtained by setting $U=0$ and
$\gamma=1$.

Within the two-site approximation, appreciable sum-rule violations
occur for $U\gg\Uco$, where the stability criterion
(\ref{RePi_omega0}) and the definition of the magnetic moment itself
constrain $\mi$ to the narrow interval between the UHF moment $\mo$
and $1$.  In this regime, the deviation of the moment from its UHF
value defines a small positive parameter $\delta=\mi-\mo$. The
position and weights of the ladder-sum poles, eqs.~(\ref{pole_rpa}),
can then be expanded in terms of the square root of this parameter
which, using $\gamma\simeq\mo\delta+\sorder{\delta^2}$, yields
\begin{subequations}
\begin{eqnarray}
  \label{poles_rpa_series}
  \label{omega_rpa_series}
  \wm^{\pm}& \simeq &
  \pm 2\wo\sqrt{\mo\delta}
  +\frac{U}{2}\mo^{2}\delta
  +\order{\sqrt{\delta}^{3}}
  \;\mbox{,}
  \\
  \label{weight_rpa_series}
  \qm^{\pm} & \simeq &
  \frac{\wo}{U\sqrt{\mo\delta}}
  \pm\frac{\mo}{2}
  +\order{\sqrt{\delta}}
  \;\mbox{.}
\end{eqnarray}
\end{subequations}
By consequence, the integrals in eq.~(\ref{Pisumrule}) evaluate to
$\qm^{+}+\qm^{-}=2\wo/U\sqrt{\mo\delta}+\sorder{\sqrt{\delta}}$ and
$\qm^{+}-\qm^{-}=\mo$; the $1/\sqrt{\delta}$-singularity of the first
result indicates a strong violation of the first sum rule (while the
second is found to be fulfilled to leading order even in this regime).

Within the present approach, this problem can be overcome by
renormalization of the ladder-sum propagator, {\sl i.e.},  by
multiplication of $\imag\Piapm(\w)$ with two different constants above
and below the Fermi level, $\w=0$, chosen to comply with both
sum rules (\ref{Pisumrule}). [The corresponding real part can
subsequently be calculated from the Hilbert transform
(\ref{Hilbert}).]  In the two-site approximation, the renormalized
weights --- which have to be used instead of eq.~(\ref{weight_rpa}) ---
are explicitly given by
\begin{equation}
\label{weight_rpa_renorm}
\qm^{\pm}=\sfrac{1}{4}(1\pm \mi)^{2}
\;\mbox{.}
\end{equation}
Henceforth, unless explicitly stated otherwise,
$\Piapm(\w)$ will always stand for such a renormalized spin-flip
ladder sum.

\subsubsection{The full spin-flip polarization propagator.}
\label{sec:fullPi}

Beyond the two-site approximation, where the polarization propagator
solely consists of two poles, with frequencies given by
eq.~(\ref{omega_rpa}) and renormalized weights
(\ref{weight_rpa_renorm}), the full ladder-sum $\Piapm$ contains the
following additional features:
 
(i) The full $\Piapm(\w)$ inherits its continua from the bare
$\Poapm(\w)$; these are located at frequencies $\w\in
\pm[\w_{\gtrless},\w_{\gtrless}+D]$ and $\abs{\w}\le 2D$.

(ii) For moments $\mi$ which only exceed the corresponding UHF moment
very slightly, as appropriate in strong coupling, $U\gg 4\Do$, the
pole frequencies (\ref{omega_rpa}) may lie within the low-energy
continuum $\abs{\w}\le 2D$; instead of poles, the full $\Piapm(\w)$
will thus have sharp resonances at these frequencies.

(iii) In addition to the above mentioned poles, the full $\Piapm(\w)$
shows a third collective pole, induced by the RPA-like structure of
eq.~(\ref{PiRPA}) and the logarithmic singularity of $\real\Poapm(\w)$ at
the upper edge of the polarization band found for
$\w\in[-\w_{<}-D,-\w_{<}]$; of tiny weight throughout the whole range
of interactions, this pole can be safely neglected in practice.

% %%%%%%%%%%%%%%%%%%%%%%%%%%%%%%%%%%%%%%%%
\subsection{The LMA self-energy}
\label{sec:selfenergy}

For a $\dn$-spin electron, the self-energy diagram
(\ref{fig:SigmaLMA}) translates to
\begin{equation}
\label{SigmaLMA}
\Sa{\dn}(\w) \,=\, U^{2}
\int\limits_{-\infty}^{+\infty} \frac{{\rm d}\w'}{2\pi\ic}\, 
\Piapm(\w')\,\Gauhf{\up}(\w-\w')
\;\mbox{;}
\end{equation}
and the $\up$-spin self-energy follows by particle-hole symmetry
$\Sa{\up}(\w)=-\Sa{\dn}(-\w)$.

In analogy to the calculus of the spin-flip polarization propagator,
useful expressions for the self-energy
$\Sa{\dn}(\w)=\ReSa{\dn}(\w)-\ic\,\sign{\w}\ImSa{\dn}(\w)$ are
obtained by inserting the Hilbert transforms for the broken-symmetry
$\Piapm$ and $\Gauhf{\up}$ in the integrands on the rhs of
eq.~(\ref{SigmaLMA}), yielding for the imaginary part
\begin{eqnarray}
\label{ImSigmaLMA}
\lefteqn{\ImSa{\dn}(\w) =  
\theta(\w) U^{2} \int\limits_{0}^{+\abs{\w}} {\rm d}\w'\,
\imag\Piapm(\w') \Dauhf{\up}(\w-\w')}
\nonumber \\ 
&&+\theta(-\w) U^{2} \int\limits_{-\abs{\w}}^{0} {\rm d}\w'\,
\imag\Piapm(\w') \Dauhf{\up}(\w-\w')
\mbox{;}
\end{eqnarray}
and a corresponding real part, $\ReSa{\dn}(\w)$, which follows by
Hilbert transform.

The evaluation of the integrals in eq.~(\ref{ImSigmaLMA}) is greatly
facilitated by the structure of the polarization propagator
$\Piapm(\w)$, whose spectral intensity resides --- as pointed out in
Sec.~\ref{sec:Pi} --- throughout the entire range of interactions
almost exclusively in two sharp low-energy modes at
$\w=\wm^{\pm}$. Except in strong coupling, $U\gg 4\Do$, the dominant
contribution to the self-energy arises by coupling these modes to the
UHF $\up$-spin orbitals, resulting in two poles or sharp resonances in
$\ImSa{\dn}(\w)$, the first occurring at $\w=\wm^{+}+{\w_{<}}>0$ (net
weight $U^2\qm^{+}q_{<}$), and the second at $\w=\wm^{-}-{\w_{>}}<0$
(net weight $U^2\qm^{-}q_{>}$).

The self-energy pole at positive frequencies is associated with
single-particle excitations where an extra $\dn$-spin electron is
introduced on the impurity, and the initial $\up$-spin impurity
electron hops off to the host. The same result could have been
obtained from the original ground-state by placing the extra electron
directly in the empty $\up$-spin UHF orbital (energy cost ${\w_{<}}$)
and simultaneously flipping the impurity moment from $\up$ to $\dn$
(energy cost $\wm^{+}$).  Analogously, the $\ImSa{\dn}$-pole at
negative frequencies is caused by a $\dn$-spin electron being taken
off the impurity and subsequently replaced by an $\up$-spin electron
from the host, yielding an $\up$-spin electron in the UHF orbital
below the Fermi level (energy gain ${\w_{>}}$) and a flip of the
impurity moment from $\dn$ to $\up$ (energy gain $\abs{\wm^{-}}$).  As
this requires the impurity to be initially $\dn$-spin occupied, the
latter self-energy contribution will decline as the impurity moment
$\mi$ approaches saturation.

In addition to these collective modes, eq.~(\ref{ImSigmaLMA}) predicts
several sets of narrow self-energy continua that arise from band
contributions in $\Dauhf{\up}$ and $\Piapm$; such continua occur for
$\w\ge\wm^{+}$, for $\w\le\wm^{-}$ and for
$\abs{\w}\ge{\w_{>}}+{\w_{<}}$ (all of width $D$), for
$\abs{\w}\ge{\w_{\gtrless}}$ (width $2D$), and for $\abs{\w}\le 3
D$. All these continua remain generally weak over the whole range of
parameters, albeit with one exception: as the impurity moment $\mi$
approaches its UHF value from above, $\mi\to\mo+0$, and consequently
$\wm^{\pm}\to 0$ --- a situation that becomes relevant in the strong
coupling Kondo limit discussed in Sec.~\ref{sec:selfenergyKondo} below
--- the minor UHF single-particle pole vanishes exponentially [see
eq.~(\ref{weight_l_lrgU})], and the dominant self-energy contribution
is transferred to the low-energy continuum located at $\abs{\w}\le 3
D$.

That the low-frequency behaviour of the latter continuum fulfils the
first condition for symmetry restoration, eq.~(\ref{LuttingerIm}), may be
seen as follows: the continuum is generated, as in eq.~(\ref{ImSigmaLMA}),
by convoluting the UHF band for $\up$-spin electrons with the
low-frequency continuum of $\imag\Piapm$; the latter vanishes linearly
in $\w$ as $\w\to 0$ [a consequence of eq.~(\ref{ImPi0_omega0}) in
combination with the analyticity condition (\ref{RePi_omega0}), which
for a ladder-sum propagator implies $1-U\real\Poapm(\w=0)>0$],
entailing their convolution to behave as
$[\ImfullSa{\dn}(\w)\equiv]\ImSa{\dn}(\w)\propto\w^{2}$ for
$\abs{\w}\ll D$.

The second condition, eq.~(\ref{LuttingerRe}), by contrast, is not
automatically fulfilled by the present theory. As pointed out before,
it reduces under particle-hole symmetry to
$\RefullSa{\dn}(\w=0)=\Sa{\dn}^{\rm stat}+\ReSa{\dn}(\w)=0$, which,
under the assumption $\Sa{\dn}^{\rm stat}\simeq U\mu/2$, becomes
\begin{equation} 
  \label{pinning}
  \ReSa{\dn}(\w=0)\,+\,\frac{U}{2}\mi\,=\,0
  \;\mbox{.}
\end{equation}

Within the LMA, this condition will be satisfied by tuning the
impurity moment $\mi$ within the interval $\mo<\mi\le 1$, prescribed by
the analyticity condition (\ref{RePi_omega0}). Despite the generally
narrow range of possible values, a solution can always be found since
the dynamical contribution, $\ReSa{\dn}(\w=0)$, is very sensitive to
the exact position of the majority self-energy pole and hence
implicitly to the spin-flip scale $\wm^{+}$. The latter quantity is
highly responsive to changes of the impurity moment and thus effectively
controls the condition (\ref{pinning}). In a more general perspective,
$\wm^{+}$ --- which vanishes for an ordinary UHF ground-state with
finite impurity moment (see Sec.~\ref{sec:RPAstability}) --- may be
considered as a new order parameter that, for a state with local
moments, determines whether Kondo physics takes place or not.

The above analysis is corroborated by the numerical results displayed
in Fig.~\ref{fig:uvsmu}, confirming that, for a wide range of
interactions, $U\gg\Uco$, the impurity moment $\mi$ needed to comply
with condition (\ref{pinning}) is indeed very close to the UHF moment
$\mo$.

In the remaining two paragraphs of this section,
condition~(\ref{pinning}) in combination with the LMA self-energy will
be studied in two regimes for which analytic results can be worked
out: (i) the two-site limit, appropriate for weak to moderate
interactions; and (ii) the strong coupling Kondo limit.

\subsubsection{The self-energy in the two-site limit.}
\label{sec:selfenergy2site}

The two-site approximation --- $D\to 0$ albeit with finite
$\wo\simeq\sqrt{2\Do D/\pi}$ --- allows for the LMA self-energy to be
written in simple analytical terms and, more importantly, suggests a
two-pole structure for $\Sa{\sigma}(\w)$ that, for $\w\gg\abs{D}$ and
up to moderate interactions $U\lesssim\Do$, correctly reproduces the
main results of a complete version of the LMA for the present
narrow-band AIM.

In weak coupling, $U\ll\Uco$, UHF predicts a solution with a vanishing
impurity moment, $\mo=0$, and single-particle propagators which,
independent of spin and impurity type, reduce to the noninteracting
Green function, $\Gasuhf(\w)\equiv\GiF(\w)$. The corresponding
self-energies $\Sa{\sigma}(\w)$ are hence odd functions of $\w$ and
their real parts satisfy the condition (\ref{pinning}) by symmetry. As
obvious from eq.~(\ref{fig:SigmaLMA}), the leading contribution to the
self-energy diagrams stems from ordinary 2PT about the noninteracting
ground state and is hence overwhelmingly dominated by poles at
$\w\simeq\pm 3\wo$ of weight $Q\sim U^{2}/8$ each.

Above some critical interaction $\Uc$ --- which is of the same order
of magnitude but slightly smaller than the corresponding UHF critical
interaction $\Uco$ --- the nonmagnetic solution becomes unstable and
a finite moment forms on the impurity. As illustrated in
Fig.~\ref{fig:uvsmu}, the LMA moment $\mi$ saturates rapidly with
increasing $U$, and its numerical values fit accurately to a
square-root law [which, in contrast to and despite the similarity with
eq.~(\ref{muuhf_twosite}), is not even exact in the two-site limit]:
\begin{equation}
\label{mulma_fit}
\mi\,\simeq\,
\begin{cases}
\sqrt{1-\left({\Uc}/{U}\right)^{2}} & \mbox{for $U>\Uc:=2\sqrt{2}\,\wo$} \cr
0 & \mbox{for $U<\Uc$}
\end{cases}
\;\mbox{.}
\end{equation}

As pointed out above, symmetry restoration and consequently the
Fermi-liquid nature of the single-particle excitations hinge, in this
regime, almost exclusively on the energy cost $\wm$ for flipping the
impurity moment. For simplicity, in what follows, the latter quantity
will be thought of as an independent parameter, to be determined from
condition (\ref{pinning}), and the impurity moment will be kept at its
UHF value $\mo$ instead.

Under these assumptions (which are in concord to leading-order with
the analysis in Sec.~\ref{sec:sumrule}), $\Piapm$ is dominated by
poles at $\wm^{\pm}=\pm\wm$, of renormalized weights
$\qm^{\pm}=\frac{1}{4}(1\pm\mo)^2$, and $\Sa{\dn}(\w)$ is constituted
by a majority pole at $\w=\wm+{\w_{<}}$ (net weight
$U^2\qm^{+}q_{<}\equiv 4\wo^{2}q_{>}$) and a minority pole at
$\w=-\wm-{\w_{>}}$ (net weight $U^2\qm^{-}q_{>}\equiv 4\wo^{2}q_{<}$).
The magnetic energy scale follows finally from eq.~(\ref{pinning}),
viz.
\begin{equation}
\label{wm2site}
\wm\,=\,-\frac{U}{4}+J\,+\,\sqrt{\left(\frac{U}{4}\right)^{2}+J^{2}+\wo^{2}}
\;\mbox{,}
\end{equation}
where, again, $J\equiv 4\wo^2/U$ stands for the antiferromagnetic
exchange coupling constant.

At the critical interaction for moment formation in UHF, where
$U=\Uco$ and $U/4=J=\wo$, the magnetic energy starts out at
$\wm=\sqrt{3}\wo$, and then decreases as $\wm\sim 3J/2$ for
interactions $U\gg\Uco$ (the relevance of the two-site approximation
being subject to $\wm\gg D$, or equivalently $U\lesssim\Do$ as
previously).

Due to the simplicity of the two-site approximation, two further
virtues of the LMA become manifest: first, the renormalization
procedure of $\Piapm(\w)$ --- which enforces the sum rules
(\ref{Pisumrule}) --- guarantees the LMA self-energy (\ref{SigmaLMA})
to fulfil an analogous set of sum rules for the interaction
self-energy $\fullSa{\dn}(\w)\equiv
U^2\GF{c_{i\dn}\delta\hat{n}_{i\up}}{\delta\hat{n}_{i\up}c^{+}_{i\dn}}_{\w}$,
\begin{eqnarray}
\label{Sigmasumrule}
\left[ \int\limits_{0}^{+\infty} \!\pm\! \int\limits_{-\infty}^{0} \right]
\frac{{\rm d}\w}{\pi} \,\ImfullSa{\dn}(\w)
& = &
U^2\mean{\comm{c_{i\dn}\delta\hat{n}_{i\up}}{\delta\hat{n}_{i\up}c^{+}_{i\dn}}_{\pm}}
\nonumber \\
& \simeq & 
\frac{U^{2}}{4}\left(1-\mo^{2}\right)\times\begin{cases} 1 & \cr \mo & \end{cases}
\,\mbox{,}
\nonumber \\
&&
\end{eqnarray}
the sole condition being that the many-body expectation values on the
rhs are evaluated by a Hartree-Fock factorization, as shown on the
second line of eq.(\ref{Sigmasumrule}) (which, again, correctly
captures the limits of weak and strong coupling).

The second virtue concerns the conventional {\em single} self-energy
$\Si{}$: in Lanczos calculations by Hofstetter and Kehrein,
\cite{Hofstetter99} this $\Si{}(\w)$ is found to have poles on an
energy scale $\w\sim\sqrt{\Do D}$ which cannot be explained in {\em
  any} order of the skeleton expansion.  \cite{Hofstetter99,Kehrein98}
Moreover, these poles have been shown to depend little or not at all
on the interaction strength, \cite{SL01} occurring in the limits of
strong {\em and} weak coupling exactly at $\w=\pm 3\wo$, with net
spectral weight $Q\sim U^2/8$ each. Responsible in the atomic limit
$V_{i\vk}\sim\wo=0$ for the $\Si{}(\w)\sim 1/\w$ characteristics of
the insulator, they are an intrinsic feature of the narrow-band
AIM.\cite{Galpin08a}

It has been argued in the beginning of this paragraph that, in weak
coupling ($U\leq\Uc$), both LMA self-energies coincide and inherit
their main properties from 2PT, whose poles in turn precisely respect
the desired properties. But also for $U\gg\Uc$, where the impurity
moment saturates and $\wm\simeq\frac{3}{2} J$, such poles occur on a
similar energy scale in the symmetry-restored single LMA self-energy,
$\Si{}(\w)$. The latter depends, as in eq.~(\ref{singleSigma}), on the
self-energies $\fullSa{\sigma}$ which in the limit $U\gg\Uc$ reduce to
\begin{equation}
\label{SigmaLMAlrgU}
\fullSa{\dn}(\w)\,\simeq\,
\frac{\frac{U}{2}\w\left[\w+\frac{U}{2}\right]}{\w^2+\frac{U}{2}\w-4\wo^2}
\end{equation}
and $\fullSa{\up}(\w)=-\fullSa{\dn}(-\w)$; insertion of which into
eq.~(\ref{singleSigma}) generates a single self-energy constituted, to
leading order in $J\sim 1/U$, by poles at $\w\simeq\pm\sqrt{5}\wo$ of
spectral weight $Q\sim U^2/8$ each.  Hence apart from the pole
frequency prefactor ($\sqrt{5}\simeq 2.23$ instead of $3$), the LMA
self-energy matches the above mentioned properties of the exact
solution also for $U\gg\Uc$.  The discrepancy in the prefactor is
possibly due to the inability of the two-site approximation to
describe the true strong coupling regime, $U\gg 4\Do$.

\subsubsection{The self-energy in the Kondo limit.} 
\label{sec:selfenergyKondo}

In genuinely strong coupling, $U\gg 4\Do$, the polarization propagator
$\Piapm(\w)$ consists in essence of a single sharp resonance peaked at
a frequency $\wm$ which is tiny compared to any other frequency scale
involved in the problem, even $D$.  The corresponding self-energy,
obtained by convoluting this resonance with the single-particle UHF
propagator as in eq.~(\ref{ImSigmaLMA}), is thus governed by resonant
spin-flip scattering within the metallic single-particle band at the
Fermi level itself, {\sl i.e.},  the Kondo effect, whereas the orbital
contributions, predominant in the two-site limit, vanish exponentially
[see eq.~(\ref{weight_l_lrgU})].

Under these circumstances, and granted $\mo\simeq 1$, condition
(\ref{pinning}) can again be solved analytically, yielding an
exponentially small spin-flip energy, $\wm\simeq D\exp[-\pi U/8\Do]$,
which --- apart from the prefactor --- concurs with the LMA results
for the AIM with a flat and infinite wide hybridization band.
\cite{Logan98} Far from the band edges of the low-energy continuum,
the interaction self-energy can be written as a function of a single
parameter, $\wtil=\w/\wm$, viz.
\begin{equation}
\label{SigmaKondo}
\fullSa{\dn}(\w)\,\stackrel{\abs{\w}\ll D}{\sim}
4\Do \left[ \frac{1}{\pi}\ln\abs{1-\wtil} - \ic\theta(\wtil-1)
\right]
\;\mbox{,}
\end{equation}
and is otherwise independent of the original parameters $U$ and $D$.

This low-fequency scaling behaviour of the self-energy -- which is
ultimately responsible for the scaling properties of the
single-particle resonance -- is a hallmark of AIMs with metallic
hosts. \cite{Logan98,Dickens01,Glossop02} The associated prediction of
an exponentially small magnetic energy scale, closely related to the
Kondo temperature, emerges similarly from Anderson's {\em poor man's
  scaling} \cite{Anderson70} for the {\it s-d} model, and from a
strong-coupling expansion of the exact Bethe ansatz solution for the
AIM. \cite{Tsvelik83}

% %%%%%%%%%%%%%%%%%%%%%%%%%%%%%%%%%%%%%%%%
\section{Spectral evolution in the narrow-band regime.}
\label{sec:spectra}

In this section, the numerically obtained LMA impurity spectra
$\Di{}(\w)=-\frac{1}{\pi}\sign{\w}\imag\Gi{}(\w)$ [with $\Gi{}(\w)$
the spin-symmetric single-particle Green function of
eq.~(\ref{Gmixspin})] will be presented next to corresponding UHF and
2PT results. The section also comprises a comparison of the LMA
spectra with Lanczos calculations performed by Hofstetter and Kehrein
\cite{Hofstetter99} for an $11+1$-site Anderson star with host
bandwidth $D=10^{-4}\Do$.

Further analytic rationales reveal how the spectrum evolves with
interaction strength and suggest an interpretation of its main
features in simple physical terms: for weak to moderate interactions,
$U\lesssim\Do$, the reasoning is based on the two-site approximation
where the host band is taken to be infinitely narrow; conversely, in
strong coupling, $U\gg 4\Do$, similarities with the opposite case of
an infinitely wide host band emerge in the immediate vicinity of the
Fermi level, and ultimately lead to the characteristic scaling
behaviour of the Kondo resonance.\cite{Logan98,Dickens01} 

\subsection{Weak coupling: \texorpdfstring{$U\ll\Uc$}{U<<Uc}}
\label{sec:spectra_lowU}

\begin{figure}[htbp]
  \begin{center} 
  \leavevmode 
  \includegraphics[width=\FigureWidth]{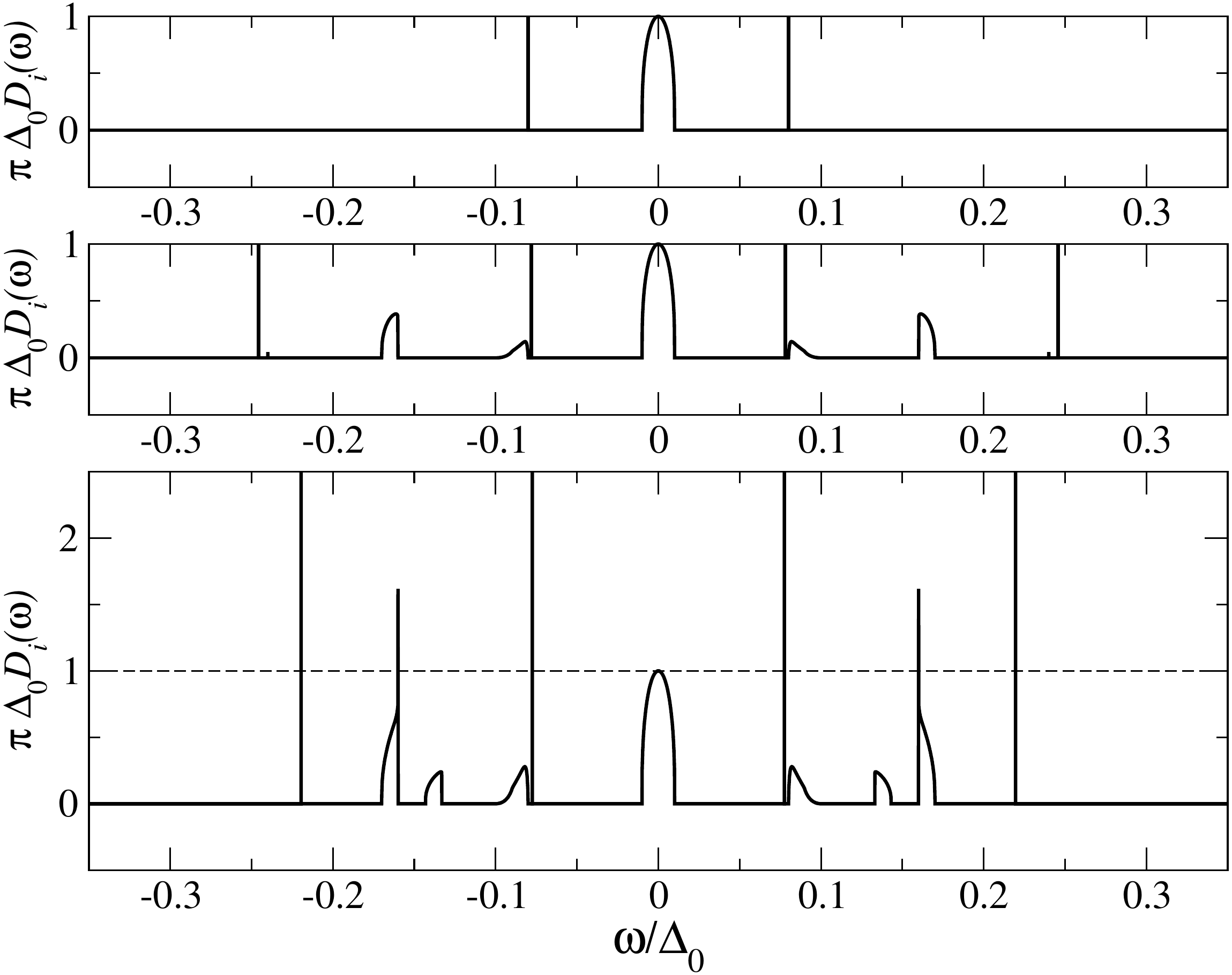}

  \caption{UHF (top), 2PT (middle), and LMA (bottom) impurity spectra,
    $\pi\Do\Di(\w)$ vs.~$\w/\Do$, for bandwidth $D=0.01\Do$, {\sl
      i.e.}, $\wo\simeq 0.08\Do$ and $\Uc\simeq 0.23\Do$, and
    interaction strength $U=0.1\Do$. Discrete levels are represented
    by vertical lines. The Fermi energy is $\w=0$.}

  \label{fig:dosweak} 
  \end{center}
\end{figure}

For $U$ much smaller than the critical interaction for moment
formation, $\Uc$, the LMA connects smoothly to 2PT (see
Sec.~\ref{sec:selfenergy2site}), which in turn exactly captures the
two-site limit for {\em any} interaction strength.  By analogy to the
latter, extensively discussed in Ref.~\onlinecite{SL01}, the LMA
single-particle spectrum (depicted in the last graph of
Fig.~\ref{fig:dosweak}) is overwhelmingly dominated by the ``molecular
orbitals,'' occurring at minimally lower frequencies
$\abs{\w}\lesssim\wo$ than in the noninteracting limit.
Additionally, for any nonzero $U$, a pair of weak poles arises at
$\w\simeq \pm 3\wo$: these are the precursors of the Hubbard
satellites, and start out with net spectral weight
$q\sim\order{[U/\wo]^{2}}$ each.

Fig.~\ref{fig:dosweak} depicts the impurity spectra for $D=0.01\Do$
and $U=0.1\Do$, {\sl i.e.}, an interaction strength that is weak
compared to $\Do$, but already appreciable with respect to $\Uc\simeq
2\sqrt{2}\wo\simeq 0.23\Do$ (or $\Uco\equiv 4\wo\simeq 0.32\Do$).
Since $U<\Uc$, the LMA converges to a solution without magnetic
moment, and the numerically computed LMA spectra (bottom panel) carry
mainly the signature of 2PT (middle panel), namely: (i) the
``molecular orbitals'', occurring at $\w\simeq \pm 0.077\Do$, dominate
the spectrum and carry in total more than $92\%$ of the intensity;
(ii) the outer set of poles, at $\w\simeq \pm 0.22\Do$, carrying
approximately $6.6\%$ of the total spectral weight, occur slightly
inside their 2PT counterparts at $\w\simeq\pm 0.25\Do$; (iii) various
weak band contributions which (with one exception) also bear great
similarity with 2PT, the most prominent being the essentially
unrenormalized Fermi-liquid continuum of width $2 D$ and net spectral
weight $\order{D/\Do}$ around the Fermi level. These results clearly
contrast with UHF, in the top panel of Fig.~\ref{fig:dosweak}, whose
spectra, for the reasons already pointed out in Sec.~\ref{sec:uhf},
coincide with the noninteracting limit.

\subsection{Moderate coupling: \texorpdfstring{$\Uc\ll U\lesssim\Do$}{Uc<<U<=Delta0}}
\label{sec:spectra_medU}

In the regime of moderate interaction strengths --- defined by $\Uc\ll
U\lesssim\Do$ or, equivalently, $D\ll J\ll\wo$ --- a well-established
local moment resides on the impurity. The energy cost
$\wm\simeq\frac{3}{2} J$ for flipping this moment being large compared
to the host bandwidth $D$, the two-site scenario of
Sec.~\ref{sec:selfenergy2site} remains appropriate --- except, of course,
within the low-energy Fermi-liquid continuum $\abs{\w}<D$ itself.

\begin{figure}[htbp]
  \begin{center} 
  \leavevmode 
  \includegraphics[width=\FigureWidth]{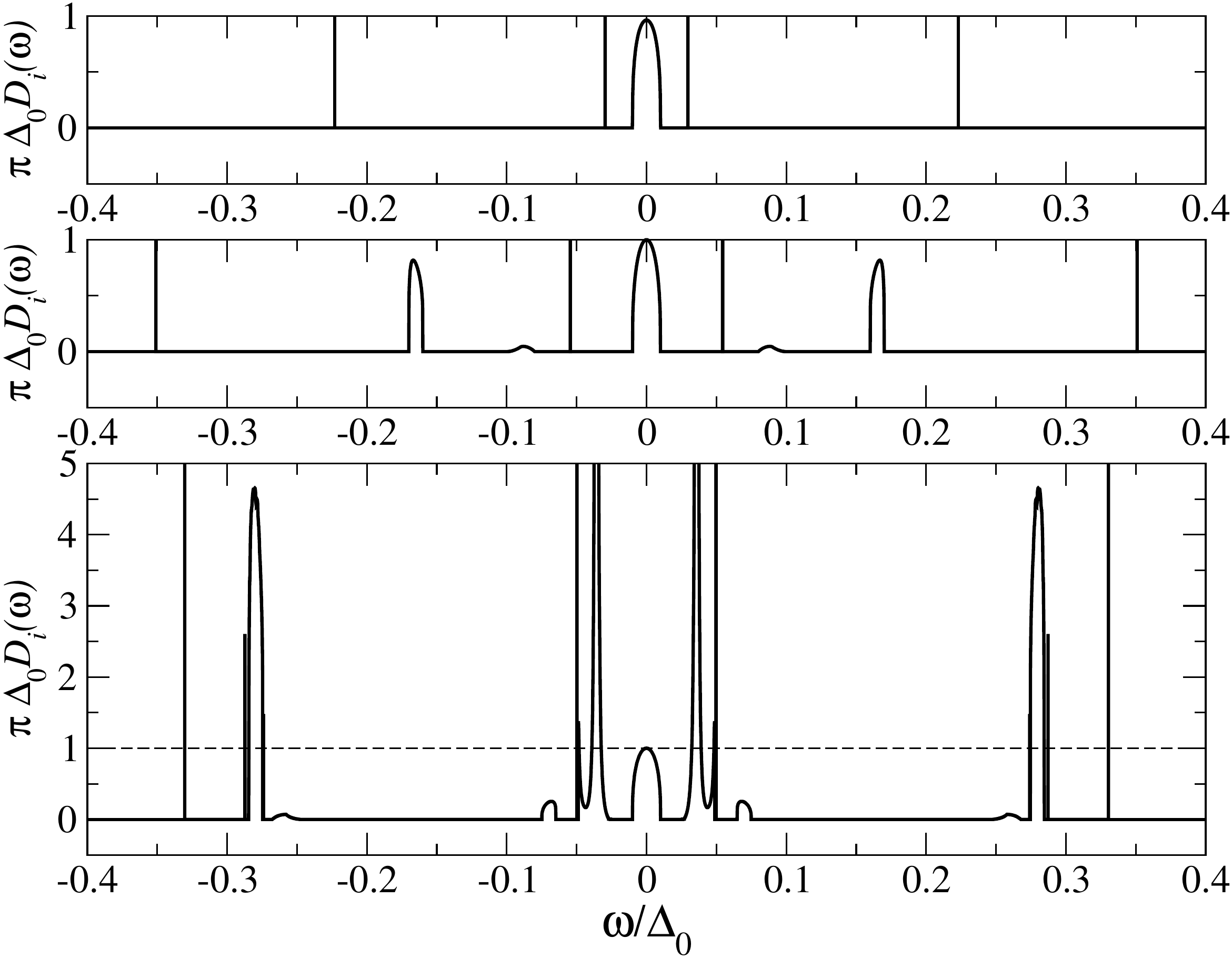}

  \caption{UHF (top), 2PT (middle), and LMA (bottom) impurity spectra,
    $\pi\Do\Di(\w)$ vs.~$\w/\Do$, for bandwidth $D=0.01\Do$, {\sl
      i.e.}, $\wo\simeq 0.08\Do$ and $\Uc\simeq 0.23\Do$, and
    interaction strength $U=0.5\Do$ ($J\simeq 0.05\Do$). Discrete
    levels are represented by vertical lines. The Fermi energy is
    $\w=0$.}

  \label{fig:dosmoderate} 
  \end{center}
\end{figure}

For a host of bandwidth $D=0.01\Do$, chosen in
Fig.~\ref{fig:dosmoderate} for clear resolution on all relevant energy
scales, the above conditions of moderate interaction strengths are
never truly met, since $\Uc\simeq 0.23\Do$ and $\Do$ are of similar
orders of magnitude. Nevertheless, for $U=0.5\Do$ (as in the figure),
where the LMA impurity moment is well established ($\mi\simeq 0.89$
vs. $\mo\simeq 0.78$ for UHF, see Fig.~\ref{fig:uvsmu}), the moderate
coupling regime of the two-site approximation is a good starting point
for the analysis of the impurity spectra.
Their most prominent feature, accounting for approximately $68\%$ of
the spectral intensity, are the high-energy Hubbard satellites at
$\abs{\w}\simeq\frac{1}{2}U+\frac{3}{2}J\simeq 0.33\Do$. They are
found to occur remarkably close to their 2PT counterparts at
$\abs{\w}\simeq 0.35\Do$ (expected in the two-site limit of 2PT at
$\abs{\w}\simeq \frac{1}{2}U+\frac{5}{2}J\simeq 0.38\Do$), whereas the
UHF correction to the atomic limit (where $\abs{\w}\equiv\frac{U}{2}$)
operates in the opposite direction, $\abs{\w}\simeq
\frac{1}{2}U-\frac{1}{2}J\simeq 0.22\Do$. A new feature on the high
energy scale are the side bands for
$\abs{\w}\ge{\w_{>}}+{\w_{<}}\simeq 0.27\Do$ (and, very faint, for
$\abs{\w}\ge\w_{>}\simeq 0.25\Do$), stemming from many-body
self-energy continua (see Sec.~\ref{sec:selfenergy}): strongly enhanced
with respect to their 2PT progenitors (located in the middle panel at
$\abs{\w}=2\wo\simeq 0.16\Do$ and $\wo\simeq 0.08\Do$), these side
bands will, for larger $U$, eventually merge with the Hubbard levels,
thereby destroying their discrete nature. This, in turn, may be viewed
as an extreme version of the many-body broadening effect generic to
the AIM and likewise observed for hosts of large
bandwidth. \cite{Logan98,Glossop02}

In the low-energy sector --- governed by the antiferromagnetic scale
$J=4\wo^2/U\simeq 0.05\Do$ and the bandwidth $D=0.01\Do$ --- the LMA
impurity spectra reveal a much richer structure than their UHF or 2PT
counterparts. While all three approaches predict the low-energy
continuum, ranging from $\w=-D$ to $D$ and accounting for $0.4\%$ of
the total spectral intensity, to suffer very little renormalization
with respect to the noninteracting limit, the spectra differ
significantly on the $J$-scale, where simpler UHF and 2PT both predict
a {\em single} pair of discrete levels (which can be viewed as
remnants of the molecular orbitals). An additional, second pair of
levels emerges in the LMA already in the framework of the two-site
approximation: the pole equation for the $\dn$-spin electron,
$\w-\ReDelta(\w)-\Sa{\dn}(\w)=0$, with $\Sa{\dn}(\w)$ from
eq.~(\ref{SigmaLMAlrgU}), yields the upper Hubbard satellite together
with {\em two} low-energy poles at, to leading order in $J$,
\begin{equation}
\label{omega_lowfreqpoles}
\w_{1,2}\simeq -\frac{1\pm\sqrt{17}}{4} \,J
\;\mbox{,}
\end{equation}
while, the $\up$-spin pole equation, leads to the lower Hubbard
satellite and poles at $\w=-\w_{1,2}$.

The low-energy poles at $\w=\pm\w_{1,2}$ can be interpreted in the
following way: the outer levels at $\abs{\w}=\abs{\w_{1}}\simeq 1.28
J\simeq 0.065\Do$ (of net weight $q_{1}\simeq 0.51[J/\wo]^{2}$),
result from a shift of the UHF orbital remnants, initially found at
$\abs{\w}=\w_{<}\simeq \frac{1}{2}J$, and occur remarkably close to
their 2PT counterparts at $\abs{\w}\simeq\frac{3}{2}J\simeq 0.076\Do$;
the inner poles, at frequencies $\abs{\w}=\w_{2}\simeq 0.78 J\simeq
0.04\Do$ and of net weight $q_{2}\simeq 0.12[J/\wo]^{2}$, are entirely
new features, which owe their existence to resonant spin-flip
scattering as accounted for by the majority self-energy pole at
$\w=\w_{<}+\wm\simeq 2J$. In UHF, this pole is naturally absent due to
the total lack of dynamics; but it also misses in 2PT, albeit for
subtler reasons: assuming a perfect parity between host and impurity,
as appropriate for an actual two-site problem, a resonance process
that tends to localize a spin flip on the impurity rather than on the
host is ruled out in 2PT by construction.

Despite the rather poor concord with the definition of moderate
coupling, the numerically calculated LMA spectrum of
Fig.~\ref{fig:dosmoderate} corroborates the above analysis to
reasonable accuracy, displaying in the low-energy sector a four-peak
structure with (i) levels at $\abs{\w}\simeq 0.05\Do\simeq J$ (instead
of the expected $1.28 J$) and carrying about $21\%$ of the spectral
intensity; and (ii) sharply peaked continua at $\abs{\w}\simeq 0.7
J\simeq 0.035\Do$, accounting for roughly $8\%$ of the spectral
weight, stemming from the merger of the above mentioned inner poles,
expected at $\w\simeq\pm 0.78 J$, with self-energy bands.

\subsection{Comparison with Lanczos spectra.}
\label{sec:spectra_lanczos}

In their article, \cite{Hofstetter99} Hofstetter and Kehrein present
low-frequency Lanczos spectra for an Anderson impurity coupled to a
host of $11$ sites, with a bandwidth of $D=10^{-4}\Do$, entailing a
``bonding energy'' of $\wo\simeq 0.008\Do\equiv 80 D$. They study two different
interaction strengths: the first, $U=0.2\Do$, implies $J\simeq 12.7 D$
much larger than $D$ but much smaller than $\Uc\simeq 230 D$, thus
matching the above definition of moderate coupling; the second,
$U=4\Do$, implying $J\simeq 0.63 D$ slightly smaller than the
bandwidth, is in the crossover region between moderate coupling and
the Kondo regime.
\begin{figure}[htbp]
  \begin{center} 
  \leavevmode 
  \includegraphics[width=\FigureWidth]{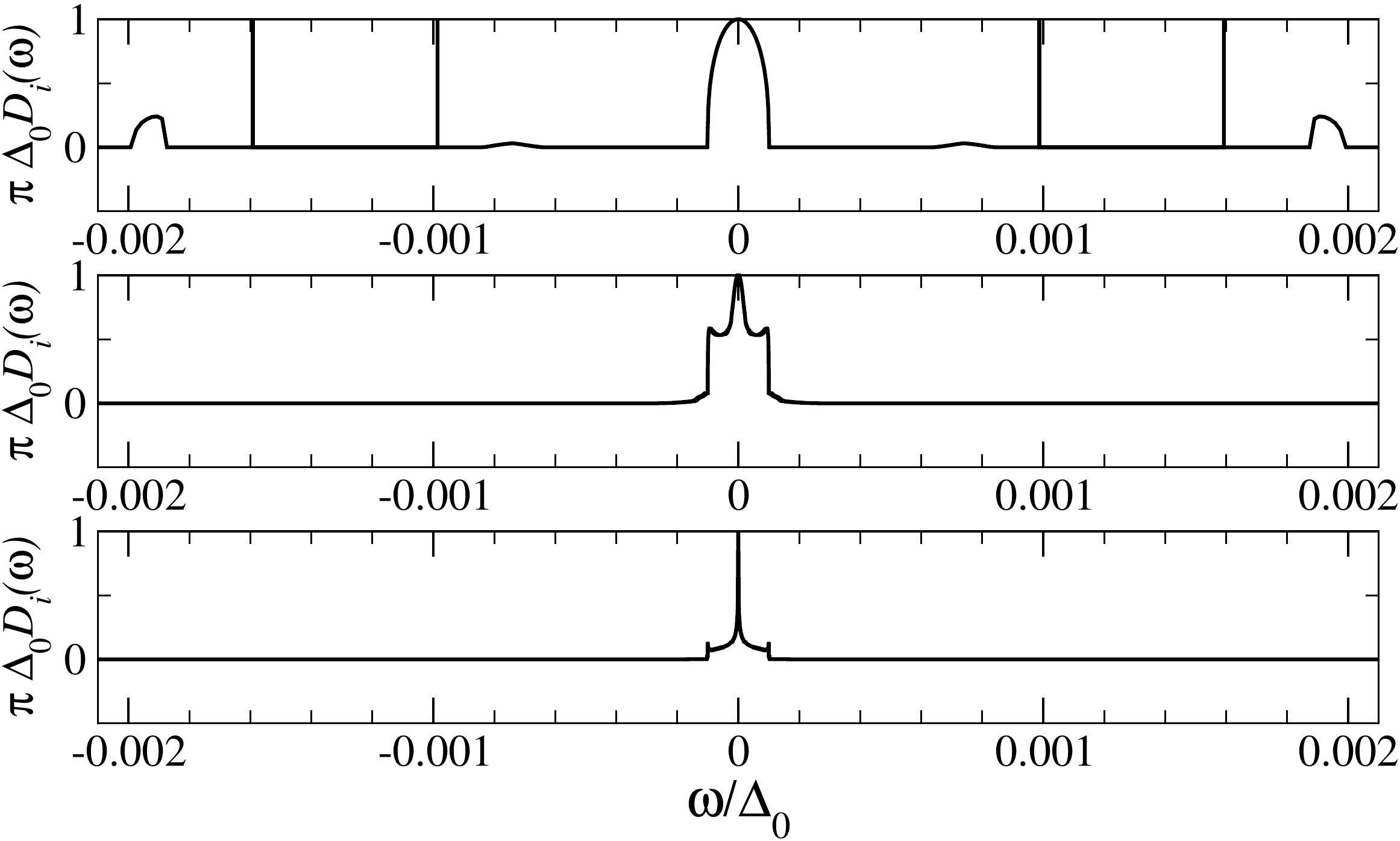}

  \caption{Low-frequency sector of the LMA impurity spectra,
    $\pi\Do\Di(\w)$ vs.~$\w/\Do$, for bandwidth $D=10^{-4}\Do$, {\sl
      i.e.}, $\wo\simeq 0.008\Do=80 D$, and interaction strengths
    $U=0.2\Do$ (top), $U=4\Do$ (middle), and $U=10\Do$ (bottom).
    Discrete levels are represented by vertical lines. The Hubbard
    satellites at $\abs{\w}\simeq U/2$ lie outside the plot range.}

  \label{fig:lmavslanczos} 
  \end{center}
\end{figure}
Fig.~\ref{fig:lmavslanczos} shows LMA spectra for the same
parameters, along with an additional third calculation for $U=10\Do$
--- which is in the true strong coupling regime difficult to reach for
the Lanczos method. 

Hofstetter and Kehrein's Lanczos spectrum for $U=0.2\Do$ is dominated
by Hubbard satellites at energies as high as
$\abs{\w}\simeq\frac{U}{2}=0.1\Do\equiv 1000 D$ which are excluded
from their plots.  \cite{Hofstetter99} In the physically more relevant
low-energy sector, their spectrum may be divided in two parts: (i) a
central continuum for $\abs{\w}\leq D\equiv 10^{-4}\Do$ which in shape
is very similar to the Fermi-liquid band of the noninteracting limit;
and (ii) two sets of narrow peaks on the antiferromagnetic scale
$\order{J\simeq 12.7D}$, the first located at $\abs{\w}\simeq
0.0018-0.0021\Do$ ({\sl i.e.}, $18-21 D$) and the second at
$\abs{\w}\simeq 0.0025-0.0026\Do$ ({\sl i.e.}, $25-26 D$).  Although it
is not clear how these sets will evolve with a larger number of sites
in the Lanczos calculations, a structure with at least two features
will most likely prevail in this region.

The corresponding LMA results, displayed in the top panel of
Fig.~\ref{fig:lmavslanczos}, reproduce --- at least qualitatively ---
all characteristics of the Lanczos spectra: excellent agreement is
found on the highest energy scale (off plot-range in the graph),
governed by the Hubbard satellites at $\w\simeq\pm 0.1\Do$ with almost
$97\%$ of the total spectral intensity; good agreement is also
observed on the lowest energy scale $D$, occupied by the central
Fermi-liquid continuum, of net weight $\order{D/\Do}\sim 0.01\%$, and
essentially unrenormalized from the noninteracting-limit, since,
throughout the whole continuum, the interaction self-energies
$\fullSa{\sigma}(\w)\sim\order{U^2/\Do}$ [eq.~(\ref{SigmaLMAlrgU})]
are weak in comparison to the hybridization
$\Delta(\w)\sim\order{\Do}$, and have thus very minor influence on the
quasi-particle properties.  Nevertheless, conceptually much simpler
2PT and UHF are similarly successful on the two latter energy scales.

On the intermediate energy scale, $J\simeq 12.7 D$, by contrast, where
UHF and 2PT both predict a single set of levels (see
Sec.~\ref{sec:spectra_medU}), solely the LMA produces a rich structure
which qualitatively resembles the Lanczos spectra.
\cite{Hofstetter99} The dominant features are two sets of poles at
$\w=\pm\w_{1,2}$ [eq.~(\ref{omega_lowfreqpoles})]: the first, with $2
q_{1}=2.4\%$ of the total spectral intensity and located at
$\abs{\w}=\abs{\w_{1}}\simeq 1.28 J\simeq 0.0016\Do\equiv 16 D$,
corresponds to shifted ``orbital remnants''; the second, with $2
q_{2}=0.6\%$ of the net spectral weight and situated at
$\abs{\w}=\abs{\w_{2}}\simeq 0.78 J\simeq 0.001\Do\equiv 10 D$, is the
novel feature related to resonant spin-flip processes (which has been
discussed in the previous paragraph).
Additionally, of the various sets of single-particle bands arising
from the self-energy continua (see Sec.~\ref{sec:selfenergy}), only the
two most prominent are visible in the graph: the first, at
$\abs{\w}\ge\wm\simeq\frac{3}{2}J\simeq 0.0019\Do\equiv 19 D$ and of
width $D$, stems from spin-flipping impurity scattering of the
metallic host electrons close to the Fermi level --- a process driving
the Kondo physics in the strong-coupling limit $U\gg 4\Do$ (see
Sec.~\ref{sec:spectra_lrgU}); a second, almost imperceptible set of
continua of width $2D$ is located at $\abs{\w}\ge\w_{<}\simeq
J/2\simeq 0.000\,64\Do\equiv 6.4 D$ and involves coupling the UHF
orbitals to the low-lying spin-flip continua.

In summary, contrary to UHF and 2PT, the LMA reproduces the main
aspects of the Lanczos spectra also on the intermediate energy scale,
but somewhat underestimates the antiferromagnetic exchange $J$. A
better quantitative match would require a renormalized spin exchange
constant $J'\sim 1.6 J$, or equivalently a Coulomb coupling $U'$
screened by the same factor for the spin channel only.

For $U=4\Do$, not yet in the strong coupling regime, both the LMA and
the Lanczos-determined spectra consist only of two appreciable
contributions: the Hubbard satellites at $\w\simeq \pm 2\Do$ on
the high energy scale, and, on the other extreme, the low-energy
Fermi-liquid continuum of width $\order{D}$ surrounding the Fermi
level, $\w=0$. The latter contribution, with its emergent central
Kondo resonance, is plotted for the LMA in the middle panel of
Fig.~\ref{fig:lmavslanczos}. The corresponding Lanczos graph by
Hofstetter and Kehrein confirms this scenario, albeit with a slightly
less developed Kondo resonance, and sharper peaks at the band edges
$\w=\pm D$.  The overall agreement of the spectra is good, but could
again be improved by renormalizing the LMA spin exchange, even though
the mismatch might also partly stem from the artificial Lorentzian
broadening or the small number of sites used in the Lanczos
calculations.

Finally, the Fermi-liquid continuum for $U=10\Do$, plotted in the last
panel of Fig.~\ref{fig:lmavslanczos}, illustrates the exponential
narrowing of the Kondo resonance with interaction strength which is
one of the hallmarks of the strong coupling regime to be discussed in
the next section.

\subsection{Strong coupling: \texorpdfstring{$U\gg 4\Do$}{U>> 4 Delta0}}
\label{sec:spectra_lrgU}

In an AIM with a narrow metallic host, the single-particle spectra
suggest the following phenomenological definition of the strong
coupling regime: an interaction strength belongs to the latter if (i)
the metallic band at the Fermi level is dominated by a narrow central
resonance whose shape is almost identical to Kondo resonances
belonging to even larger interaction strengths; (ii) the Hubbard
satellites appear as sharply peaked continua instead of discrete
levels; and (iii) the spectrum contains no other visible features.

\begin{figure}[htbp]
  \begin{center} 
  \leavevmode 
  \includegraphics[width=\FigureWidth]{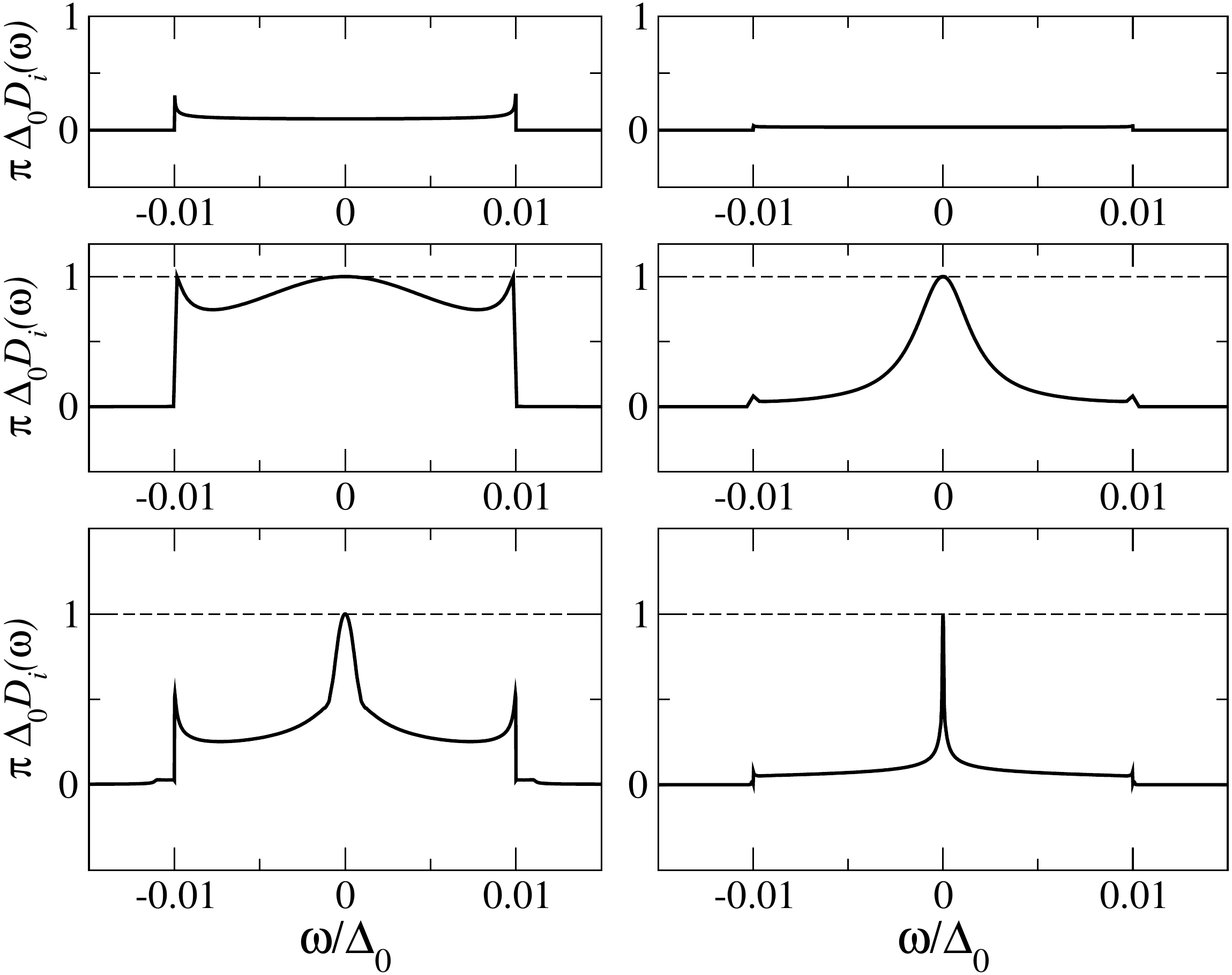}

  \caption{Low-frequency impurity spectra, $\pi\Do\Di(\w)$
    vs.~$\w/\Do$, for UHF (top), 2PT (middle), and LMA (bottom), for
    bandwidth $D=0.01\Do$ and interaction strengths $U=6\Do$ (left
    column) and $U=12\Do$ (right column). The high-energy Hubbard
    satellites are far off plot range. }

  \label{fig:dosstrong} 
  \end{center}
\end{figure}
For $U=6\Do$ and $12\Do$, the LMA single-particle spectra do indeed
comply with the last two of the above conditions. Whether however the
shape of the Kondo resonance in the low-frequency sector of the
spectra --- displayed in the bottom row of Fig.~\ref{fig:dosstrong}
--- is already scaling invariant is less obvious, but can be worked
out in analogy to the AIM with an infinitely wide metallic host:
\cite{Logan98,Dickens01} for $U\gg 4\Do$, and far from the band edges,
$\abs{\w}\ll D$, the contributions to
$\Ga{\sigma}(\w)=[\w-\Delta(\w)-\fullSa{\sigma}(\w)]^{-1}$ of both,
$\w$ and $\ReDelta(\w)\simeq 2\Do\w/\pi D$, are negligible compared to
the remaining two terms, given by $\ImDelta(\w)\simeq\Do$, and the
interaction self-energy $\fullSa{\sigma}(\w)$. In the frequency range
considered here, the latter is proportional to $\Do$ and scales in
terms of the single variable $\wtil\equiv\w/\wm$, with $\wm(U)\simeq
D\exp[-\pi U/8\Do]$ [see eq.~(\ref{SigmaKondo})]. This, in turn, leads
to scaling invariance for the part of the single-particle continuum
closest to the Fermi level, $\abs{\w}\ll D$, {\sl i.e.},  for the
central Kondo (or Abrikosov-Suhl) resonance.
\begin{figure}[htbp]
  \begin{center} 
  \leavevmode 
  \includegraphics[width=\FigureWidth]{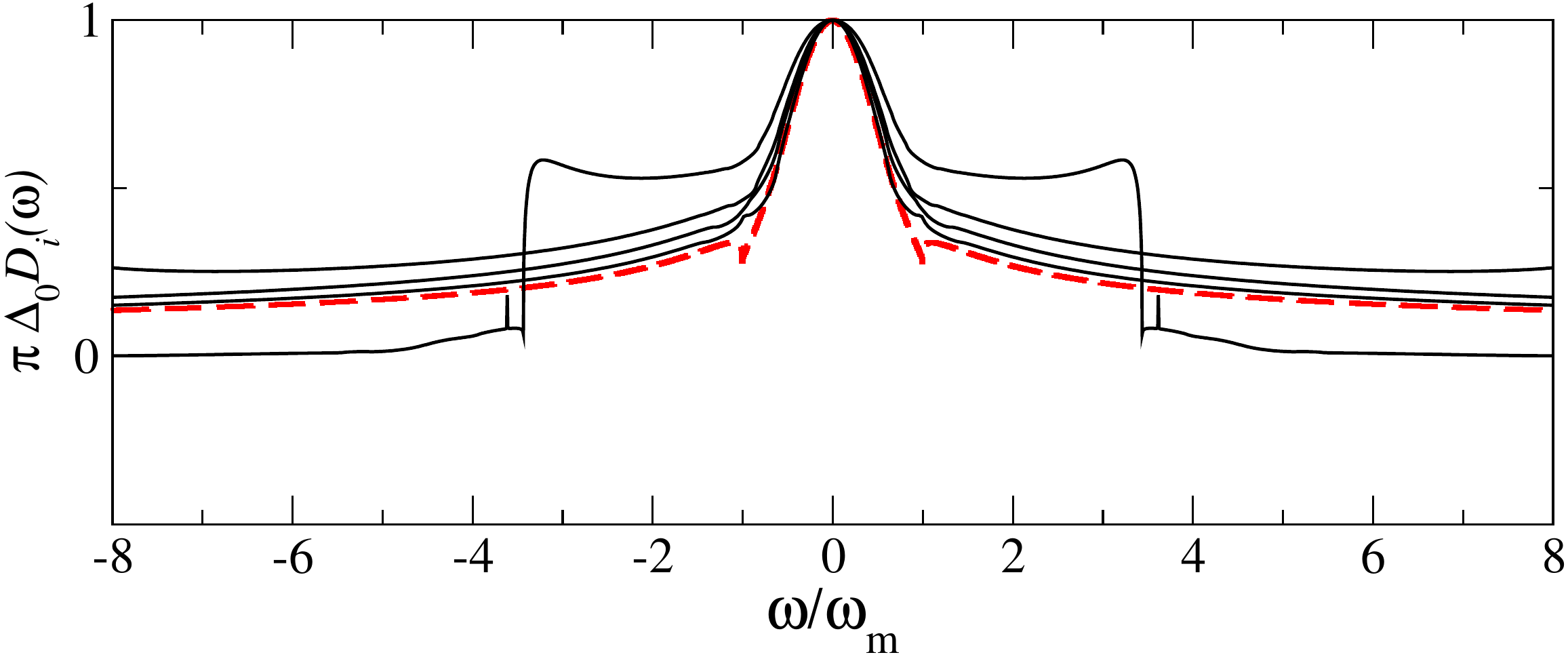}

  \caption{Scaling of the Kondo resonance in the LMA impurity spectra,
    $\pi\Do\Di(\w)$ as a function of $\w/\wm$, with $\wm$ the
    ($U$-dependent) spin-flip or Kondo energy, for bandwidth
    $D=0.01\Do$ and interaction strengths $U/\Do=4$, $6$, $8$, $12$
    (solid lines from top to bottom), and in the limit
    $U/\Do\to\infty$ (red dashed line). Full discussion in text.}

  \label{fig:lmakondo} 
  \end{center}
\end{figure}

The limiting curve for $U/\Do\to\infty$, plotted with red dashes in
Fig.~\ref{fig:lmakondo}, is identical for {\em any} symmetric AIM with
a metallic host. It comprises two parts, the first being the narrow
central region between the cusps at $\abs{\w}\simeq\wm$, within
which the line shape mostly carries the Lorentzian signature of Landau
quasi-particles of weight
$Z=\{1-[\partial\real\Si{}(\w)/\partial\w]_{\w=0}\}^{-1}$, while the
second region, outside the cusps, is characterized by long
logarithmically decaying spectral tails.  \cite{Dickens01}
\footnote{These spectral tails have formerly been considered to be of
  Doniach-\u{S}unji\'{c} form, \cite{Doniach70,Frota86,Seridonio09}
  governed by an asymptotic algebraic falloff ($\abs{\w}^{-1/2}$),
  instead of the actual logarithmic
  behaviour. \cite{Dickens01,Glossop02,Rosch03}} 
(The cusps in the spectrum at $\abs{\w}\simeq\wm$ are artefacts
stemming from the RPA-like structure of the polarization propagator
and can be removed by a more realistic ansatz for the latter,
\cite{Logan98} producing LMA line shapes which then excellently agree
with corresponding NRG data. \cite{Dickens01})

The graphs in the first two rows of Fig.~\ref{fig:dosstrong} show
clearly that such subtle effects are out of reach for the other two
methods studied here: UHF, on the one hand, suppresses for
$U/\Do\to\infty$ all spectral intensity close to the Fermi level as a
result of an incipient transition towards an insulating atomic-limit
state; 2PT, on the other hand, correctly captures the persistent
metallic character of the system -- manifest in the Friedel sum rule
pinning of the spectra at the Fermi level -- but misses the
exponential narrowing and the non-Lorentzian shape of the Kondo
resonance.

Relatedly, on the high energy scale $\abs{\w}\sim U$, the LMA
correctly predicts sharply peaked Hubbard bands instead of genuine
levels, ensuing from the absorption of the Hubbard ``levels'' at
$\abs{\w}\simeq\frac{U}{2}+\frac{3 J}{2}$ by their former
``side bands'' at $\frac{U}{2}\lesssim\abs{\w}\lesssim\frac{U}{2}+2D$;
again, UHF and 2PT both fail to catch this many-body broadening
effect.

Finally, all methods --- UHF, 2PT, LMA, and also Lanczos --- agree in
the large spectral gap ranging from the low-energy continuum of width
$D$ to the high-energy Hubbard satellites.  Especially the orbital
remnants, which in weak and moderate coupling are found inside this
gap, are now missing. A more rigorous analysis shows that they in fact
still exist, occurring exponentially close to the band edges
$\abs{\w}=D$, but are undetectable because their spectral weight
vanishes exponentially.

% %%%%%%%%%%%%%%%%%%%%%%%%%%%%%%%%%%%%%%%%%%%%%%%%%%%%%%%%%%%%
\section{Conclusions}

In this article, a symmetric AIM with a narrow metallic host has been
studied with three different theoretical approaches: UHF, 2PT, and the
LMA.

The persistent metallic character of the system -- which renders
diagrammatic perturbation expansions (like 2PT) viable in the first
place -- entails that its dynamics hinges on the impurity Coulomb
repulsion $U$ and the noninteracting Green function $\GiF(\w)$. The
latter, in turn, depends solely on the hybridization $\Delta(\w)$,
which, for the present case of a narrow host band, is mainly
determined by its width $D$ and strength at the Fermi level $\Do$.
Despite this reduced set of ultimately three parameters -- $\Do$, $D$,
and $U$ -- the single-particle spectra are found to be rich, in
particular in the physically relevant low-energy sector close to the
Fermi surface, indicating the competition of various physical
processes and their associated energy scales. These encompass
Fermi-liquid behaviour within the low-energy continuum of width $D$,
molecular orbital formation related to the bonding energy
$\wo\sim\sqrt{\Do D}$, antiferromagnetic phenomena driven by the
exchange coupling $J=4\wo^2/U$, and finally the Kondo effect with its
magnetic scale $\wm\simeq D\exp[-\pi U/8\Do]$.

Up to moderate interaction strengths, $U\lesssim\Do$, the low-energy
physics is dominated by the {\em integrated} hopping between host and
impurity, $\wo^{2}=\frac{1}{\pi}\int{\rm d}\w\ImDelta(\w)$, and the
antiferromagnetic exchange, $J$, which are both insensitive to the
details of the hybridization function. The precise form of
$\Delta(\w)$ therefore only determines the essentially unperturbed
metallic band surrounding the Fermi level. In the opposite limit of
large interactions, $U\gg\Do$, the low-energy physics is governed by
the Kondo effect. This leads to a spectral scaling, in terms of
$\w/\wm$, which, again, does not depend on any details of the
hybridization function.  

The LMA has been shown to produce meaningful results over the entire
range of interaction strengths. For small interactions,
$U\lesssim\wo$, it predicts a vanishing impurity moment and connects
smoothly to 2PT, yielding single-particle spectra carrying the
signature of prevailing orbital physics.

For moderate coupling strengths, $\wo\ll U\lesssim\Do$ (or
alternatively $D\ll J\ll \wo$), in addition to the high-energy Hubbard
levels and the essentially unrenormalized Fermi-liquid continuum on
the lowest energy scale, the LMA produces rich spectra on the
$J$-scale: here, two pairs of poles along with several accompanying
sidebands can be observed. Similar structures were found in
corresponding Lanczos-determined spectra by Hofstetter and Kehrein.
\cite{Hofstetter99} Their main contributions can be rationalized as
follows: one pair of poles may be considered as remnants of the
molecular bonding and anti-bonding orbitals which similarly occur in
UHF and 2PT; 
the other pair -- which misses in the latter approaches and also in
other state-of-the art techniques like, e.g., slave-boson based
methods \cite{Vaugier07} -- is a somewhat unexpected feature and
arises due to resonant collective spin-flip processes between the
impurity and the UHF orbital remnants. Similar spectral contributions,
in the form of sharp resonances at the inner band edges of the Hubbard
satellites, have been observed numerically in the metallic phase close
to the Mott transition, which occurs in the infinite-dimensional
Hubbard model. \cite{Nishimoto04,Karski05,Karski08} These features
have been dubbed {\em antipolarons} by Karski {\sl et al.} who suspect
them, mainly on energetic grounds, to take their origin in
bonding/antibonding phenomena between heavy quasiparticles and
collective spin excitations. \cite{Karski05} Although the same authors
admit in a subsequent publication \cite{Karski08} that this ``complex
composite excitation is not yet understood,'' their original idea is
supported by the above interpretation of the corresponding AIM
features. Karski {\sl et al.}'s results for the Hubbard model clearly
indicate that sharp features will prevail in the presence of host or
bath correlations, although their aspect may naturally be altered,
especially if they coincide with another band.

In the limit of large interaction strengths, $U\gg\Do$, of the three
investigated methods only the LMA captures simultaneously the
many-body broadening of the high-energy Hubbard satellites and the
relevant low-energy phenomena embodied in the exponentially narrowing
Abrikosov-Suhl or Kondo resonance with its distinct logarithmically
decaying wings. \cite{Dickens01,Glossop02} Moreover, the observed
Kondo resonance has been shown to possess the universal shape and
scaling properties characteristic for symmetric AIMs with a metallic
host.

The fairly accurate analytic solutions obtained within the framework
of the LMA suggest a classification into two regimes, which cover
almost the entire range of interactions: the first englobes weak and
moderate interactions, $U\lesssim\Do$, and its Fermi-liquid properties
are primarily inherited from the noninteracting system, while its
orbital contributions follow directly from the two-site approximation
where the narrow host is treated as a single site; in marked contrast
to the latter is the second regime, suitable for large interactions,
$U\gg\Do$, where the narrow host band behaves as infinitely wide in
comparison to the exponentially small Kondo energy.

\acknowledgments

First and foremost, I would like to thank David Logan for drawing my
attention to the subject and for countless valuable discussions.  I am
also most grateful to Florian Gebhard, Matthew Glossop, Nigel Dickens,
Adam Kirrander, Roland Hayn, and Laurent Raymond for constructive
comments and remarks.
The present work was initiated at the Physical and Theoretical
Chemistry Laboratory, Oxford, whom I would like to thank for their
hospitality, and with financial support of the Deutscher Akademischer
Austauschdienst (DAAD) under Grant No.~D/98/27069.

% %%%%%%%%%%%%%%%%%%%%%%%%%%%%%%%%%%%%%%%%%%%%%%%%%%%%%%%%%%%%
% REFERENCES
% %%%%%%%%%%%%%%%%%%%%%%%%%%%%%%%%%%%%%%%%%%%%%%%%%%%%%%%%%%%%

\end{document}